\documentclass[epj]{svjour}
\usepackage{graphics}
\usepackage[usenames,dvipsnames]{xcolor}
\usepackage{graphics, inputenc}
\usepackage{graphicx,amsmath,amssymb,tabularx}
\usepackage{tikz,pgfplots}
\usepackage{hyperref}
\usepackage{multicol}
\usepackage{epstopdf}
\usepackage{float}
\usepackage{subfigure}
\usepackage{mathtools}
\usepackage{appendix}
\definecolor{dgreen}{rgb}{0,.5,0}
\definecolor{dred}{rgb}{.7,.0,.0}
\newcommand{\manu}[1]{{\textcolor{dgreen}{ Manu: #1 }} }

\DeclarePairedDelimiterX\braket[2]{\langle}{\rangle}{#1 \delimsize\vert #2}
\def\ddroit{{\rm d}}

\newcommand{\be}{\begin{eqnarray}}
\newcommand{\ee}{\end{eqnarray}}
\begin{document}
\title{
Exploring weight-dependent density-functional
approximations for ensembles in the Hubbard dimer 
%Exploring the design of weight-dependent density-functional
%approximations for ensembles in the two-site Hubbard model
}
\author{Killian Deur, Laurent Mazouin, Bruno Senjean, and Emmanuel Fromager
\thanks{
%These authors contributed equally.
Corresponding author:
\href{mailto:fromagere@unistra.fr}{fromagere@unistra.fr}
}% etc
%: K.Deur (\href{mailto:deur@unistra.fr}{deur@unistra.fr}),
%L. Mazouin (\href{mailto:mazouin@unistra.fr}{mazouin@unistra.fr})
%B. Senjean (\href{mailto:senjean@unistra.fr}{senjean@unistra.fr})
}
\institute{Laboratoire de Chimie Quantique,
Institut de Chimie, CNRS / Universit\'{e} de Strasbourg,
4 rue Blaise Pascal, 67000 Strasbourg, France}
\date{Received: date / Revised version: date}
\abstract{
Gross--Oliveira--Kohn density-functional theory (GOK-DFT) is an
extension of DFT to excited states where the basic variable is the
ensemble density, i.e. the weighted sum of ground- and excited-state
densities. The ensemble energy (i.e. the weighted sum of ground- and
excited-state energies) can be obtained variationally as a functional of
the ensemble density. Like in DFT, the key ingredient to model in GOK-DFT is the
exchange-correlation functional. Developing density-functional
approximations (DFAs) for ensembles is a complicated task as both
density and weight dependencies should in principle be reproduced. In a recent paper [Phys.
Rev. B 95, 035120 (2017)], the authors applied exact GOK-DFT to
the simple but nontrivial Hubbard dimer in order to investigate
(numerically) the
importance of weight dependence in the calculation of excitation
energies. In this work, we derive analytical DFAs for various
density and correlation regimes by means of a Legendre--Fenchel
transform formalism. Both functional and density driven errors are
evaluated for each DFA. Interestingly, when the 
ensemble exact-exchange-only functional is used, these errors can
be large, in particular if the dimer is symmetric, but they cancel each
other so that the excitation energies obtained by linear interpolation
are always accurate, even in the strongly
correlated regime. 
\PACS{
      {PACS-key}{discribing text of that key}   \and
      {PACS-key}{discribing text of that key}
     } % end of PACS codes
} %end of abstract

\authorrunning
\titlerunning 
\maketitle
\section{Introduction}
\label{sec:intro}

Even though the Gross--Oliveira--Kohn ensemble density-functional theory
(eDFT)~\cite{JPC79_Theophilou_equi-ensembles,PRA_GOK_RRprinc,PRA_GOK_EKSDFT,GOK3}
for excited states is not routinely used nowadays for the
computation of excited state properties, the approach has regained interest in
recent 
years~\cite{franck2014generalised,Burke_ensemble,yang2014exact,pernal2015excitation,yang2017direct,JCP14_Filatov_conical_inter_REKS,Filatov-2015-Wiley,filatov2015ensemble,filatov2016self,filatov2017description,gould2017hartree,gould2018charge}. 
Indeed, eDFT stands as a potential alternative to
the popular time-dependent linear response DFT for the description of
charge transfer excitations, near-degeneracies and multiple electronic
excitations. Note that, in addition to eDFT, other in-principle-exact
time-independent 
extensions of DFT to excited states 
have been explored
over the years (mostly at the formal level) by Levy and coworkers, either by considering special cases where
the standard universal functional yields excited-state
energies~\cite{PRB85_Levy_Perdew_excited_states_from_GS_theory}, or by
using the external potential (or its ground-state density) as an
additional variable, thus forming a bifunctional~\cite{PRL99_Levy_F_for_excited_state,PRA99_Levy_Nagy_Koopmans_DFT,PRA01_Levy_DFT_for_degenerate_excited_states,PRA09_Ayers-Levy-dft_for_pure_ES_unification}. The latter
complication can actually be overcome for Coulomb
systems~\cite{PRA12_Nagy_TinD-DFT_ES}. Note that all these formulations are state
specific, i.e. one specific excited-state density is targeted. This is
an important difference with eDFT where the basic variable is a
state-averaged density.\\

In eDFT, weights are
assigned to the ground and the excited states that belong to the
ensemble under study. Therefore, it is in
principle crucial to model, for a fixed density, the weight dependence of the
ensemble exchange-correlation density-functional energy. Let us stress that, in
the general formulation of the theory, the
weights only need to be ordered (the largest one being assigned to the
ground state). Boltzmann weights can of course be employed~\cite{PRA13_Pernal_srEDFT} but it is not
compulsory. Using
fixed (energy-independent) weights might indeed be appealing for practical
calculations~\cite{senjean2015linear,senjean2016combining}. 
One of the limitations of approximate eDFT is the
so-called ghost-interaction error~\cite{ensemble_ghost_interaction} 
which arises when the Hartree energy
(which is quadratic in the density)
is computed with the ensemble density (i.e. the weighted sum of ground-
and excited-state densities). This issue is actually related to the proper 
description of the weight dependence in the
exchange 
energy~\cite{pastorczak2014ensemble,alam2016ghost,alam2017combining}.\\

In order to investigate the weight dependence of both exchange and
correlation density functionals, some of the authors have recently
applied eDFT to the two-site Hubbard model~\cite{deur2017exact}.
Despite its
simplicity, the model is actually nontrivial and can be used as a lab for
testing new ideas in
DFT~\cite{carrascal2015hubbard,PRB16_Burke_thermal_DFT_dimer,fromager2015exact,senjean2017local,senjean2017site}. 
The work
presented in Ref.~\cite{deur2017exact} deals with the exact theory,
which means that exact ensemble correlation energies have been computed
numerically for various density and correlation regimes. Even though
these results are precious for parameterizing density-functional
approximations (DFAs), it is still unclear how this can actually be
achieved. Following 
Carrascal {\it et al.}~\cite{carrascal2015hubbard}, 
we propose to expand the
exact ensemble correlation functional around specific cases like, for
example, the symmetric one. A general strategy, that we expect to be
applicable not only to the Hubbard dimer but also to more realistic
models, will be presented. It uses Legendre--Fenchel
transforms~\cite{LFTransform-Lieb} instead of the more popular Levy--Lieb constrained search
formalism~\cite{levy1979universal}. The paper is organized as follows. After a brief review on
exact
eDFT and its application to the Hubbard dimer (Sec.~\ref{sec:theory}),
the derivation of Taylor expansions for the exact ensemble correlation
energy will be presented in several density and correlation regimes
(Sec.~\ref{sec:Taylor_exp_various_regimes}). Following a summary of 
simple and more advanced DFAs based on the latter expansions
(Sec.~\ref{sec:comput_details}), results obtained for the total
two-state singlet ensemble
energy and the corresponding excitation energy are presented and
discussed in Sec.~\ref{sec:Results}. Conclusions are finally given in
Sec.~\ref{sec:Conclusion}. 

\section{Theory}
\label{sec:theory}

For sake of clarity, a brief introduction to eDFT and its application to the Hubbard dimer is
given in this section. More details can be found in Ref.~\cite{deur2017exact} and the
references therein.

\subsection{Ensemble density-functional theory for excited states}\label{subsec:GOKDFT}

The ensemble energy is a convex combination of $N$-electron ground- and
excited-state energies with ordered coefficients (also called weights),
the largest one being assigned to the ground state. In the particular
case of two states (ground and first-excited) which is considered in
this work, the ensemble energy associated to the electronic Hamiltonian
with local potential $v$,
\be\label{eq:Hamilto_v}
\hat{H}[v] = \hat{T} + \hat{W}_{\rm ee} + \int \ddroit{\bf r}\;v
({\bf r})\hat{n}({\bf r}),
\ee
reads
\be\label{eq:ens_ener_ab_init}
E^w[v]=(1-w)E_0[v]+wE_1[v],
\ee
where $E_0[v]$ and $E_1[v]$ are the ground- and first-excited-state
energies of $\hat{H}[v]$, and the ensemble weight $w$ is such that
$(1-w)\geq w\geq 0$ or, equivalently,
\be
0\leq w\leq 1/2.
\ee
In Eq.~(\ref{eq:Hamilto_v}), $\hat{T}$, $\hat{W}_{\rm ee}$ and $\hat{n}({\bf r})$ denote
the kinetic energy, two-electron repulsion and density 
operators, respectively. Gross, Oliveira and Kohn have shown~\cite{PRA_GOK_EKSDFT} that the
ensemble energy is a functional of the ensemble density,
\be
n^w[v]({\bf r})=(1-w)n_0[v]({\bf r})+w\,n_1[v]({\bf r})
,
\ee 
where $n_0[v]$ and $n_1[v]$ denote the ground- and first-excited-state
densities of $\hat{H}[v]$, and that it can be determined variationally
as follows,
\be
\label{variationalprinciple_dens}
E^{w}[v] = \underset{n}{\rm inf} \left\lbrace F^{w}[n] + \int \ddroit{\bf r}\,v({\bf r}) n({\bf r})\right\rbrace,
\ee
where $F^{w}[n]$ is the $w$-dependent analog of the Hohenberg--Kohn
universal functional for ensembles. While it is usually expressed within
the Levy--Lieb constrained-search formalism~\cite{levy1979universal}, which would
involve two many-body wavefunctions, we will instead use a 
Legendre--Fenchel transform-based expression. The latter is simply
obtained from Eq.~(\ref{variationalprinciple_dens}) 
by considering a fixed density $n$ and writing, for {\it any} potential
$v$, the following inequality,   
\be
E^{w}[v]\leq F^{w}[n]+\int \ddroit{\bf r}\,v({\bf r}) n({\bf r})  
,\ee
or, equivalently,
\be
F^{w}[n]\geq E^{w}[v]-\int \ddroit{\bf r}\,v({\bf r}) n({\bf r})
,
\ee
thus leading to the final expression
\begin{eqnarray}
\label{universalfunctional}
F^{w}[n] = \underset{v}{\rm sup}\left\lbrace E^{w}[v] - \int \ddroit{\bf r}\,v({\bf r}) n({\bf r})\right\rbrace.
\end{eqnarray}
As discussed further in the rest of
this work, the latter expression has the advantage of using a single variable,
namely the local potential $v$, and will not require the use of
many-body wavefunctions, which is extremely convenient for deriving
density-functional approximations.\\

In the conventional Kohn--Sham (KS) formulation of eDFT~\cite{PRA_GOK_EKSDFT}, the universal
ensemble functional is split into the noninteracting analog of
$F^{w}[n]$, namely the noninteracting ensemble kinetic energy
functional $T^{w}_{\rm s}[n]$, and the complementary ensemble 
Hartree, exchange and correlation (Hxc) density-functional energies,
\begin{eqnarray}
\label{KSdecomposition}
F^{w}[n] = T^{w}_{\rm s}[n] + E_{\rm H}[n]+E^{w}_{\rm x}[n]+E^{w}_{\rm
c}[n].
\end{eqnarray}
In analogy with Eq.~(\ref{universalfunctional}), we have
\begin{eqnarray}
\label{eq:Tsw_ab_init}
T^{w}_{\rm s}[n]= \underset{v}{\rm sup}\left\lbrace \mathcal{E}_{\rm KS}^{w}[v] - \int \ddroit{\bf r}\,v({\bf r}) n({\bf r})\right\rbrace,
\end{eqnarray}
where $\mathcal{E}_{\rm KS}^{w}[v]$ is the ensemble energy of $\hat{T}+\int \ddroit{\bf r}\;v
({\bf r})\hat{n}({\bf r})$.  
Note that, in the decomposition of Eq.~(\ref{KSdecomposition}), the conventional (weight-independent) Hartree functional is
used,
\begin{eqnarray}\label{eq:EH_def} E_{\rm H}[n] = \frac{1}{2} \iint
\ddroit{\bf r} \ddroit{\bf r'}
~ \frac{n({\bf r})n({\bf r'})}{\mid{{\bf r}-{\bf r'}}\mid}, 
\end{eqnarray}
which, in practice, can induce substantial ghost interaction
errors~\cite{ensemble_ghost_interaction,pastorczak2014ensemble,alam2016ghost,alam2017combining}. In the exact theory, the latter are removed by the
{\it weight-dependent} exchange and correlation functionals. 
Regarding the exchange energy, a general expression has been derived
and tested recently in Refs.~\cite{gould2017hartree,gould2018charge}. It allows for
the construction of an ensemble exact exchange functional (EEXX) from
the exact density (if available) or through an
optimized
effective potential (OEP) procedure.
In the Hubbard dimer, which is studied in this
work, the EEXX energy is an explicit functional of the
density~\cite{deur2017exact}.
We will therefore focus in the following on the
weight dependence of the correlation energy.\\

According to Eqs.~(\ref{variationalprinciple_dens}) and (\ref{KSdecomposition}), for a given local external potential $v_{\rm ext}$, the exact ensemble
energy $E^{w}=E^{w}[v_{\rm ext}]$ is obtained variationally as follows in KS-eDFT,
\be\label{eq:Ew_KS-eDFT_ab_init}
E^{w}= \underset{n}{\rm inf} 
\Big\{&&T^{w}_{\rm s}[n] + E_{\rm H}[n]+E^{w}_{\rm x}[n]+E^{w}_{\rm
c}[n]
\nonumber
\\&&  + \int \ddroit{\bf r}\,v_{\rm ext}({\bf r}) n({\bf r})\Big\}.
\ee

The ensemble non-interacting kinetic energy functional is usually expressed in
terms of the KS orbitals, thus leading to the
analog for ensembles of the self-consistent KS
equations~\cite{PRA_GOK_EKSDFT}. This
step is actually unnecessary in the Hubbard dimer since the exact
analytical expression for $T^{w}_{\rm s}[n]$ is known~\cite{deur2017exact}.\\

As readily
seen from Eq.~(\ref{eq:ens_ener_ab_init}), the ensemble energy varies
linearly with the ensemble weight. Consequently, the excitation energy
(or optical gap)
$\Omega=E_1[v_{\rm ext}]-E_0[v_{\rm ext}]$ can be determined either by differentiation, 
\be\label{eq:dEw_dw_exact}
\Omega=\dfrac{\ddroit E^w}{\ddroit w},
\ee
or by linear interpolation~\cite{senjean2015linear},
\be\label{eq:LIM_exact}
\Omega=2\left(E^{w=1/2}-E^{w=0}\right).
\ee
Eqs.~(\ref{eq:dEw_dw_exact}) and (\ref{eq:LIM_exact}) are equivalent in the exact theory. However,
as clearly illustrated in the following, they will give different
results, that might also be weight-dependent, when DFAs are used, as expected~\cite{deur2017exact}.
Note that, by using the stationarity of the minimizing ensemble density
$n^w=n^w[v_{\rm ext}]$ in Eq.~(\ref{eq:Ew_KS-eDFT_ab_init}), we obtain
from Eq.~(\ref{eq:dEw_dw_exact}) the simplified in-principle-exact expression,  
\be\label{eq:XE_deriv_ab_init}
\Omega = \left[\dfrac{\partial T_{\rm s}^w[n]}{\partial
w}+  
\dfrac{\partial E_{\rm x}^w[n]}{\partial
w}
+
\dfrac{\partial E_{\rm c}^w[n]}{\partial
w}
\right]_{n=n^w}
,
\ee
where, according to Eq.~(\ref{eq:Tsw_ab_init}), the first term on the
right-hand side is nothing but the KS optical gap~\cite{PRA_GOK_EKSDFT}, and the last two terms correspond to
exchange and correlation derivative discontinuity
contributions~\cite{PRA_Levy_XE-N-N-1}.\\  

Let us finally stress that the expression for $\ddroit E^w/\ddroit w$ given in the right-hand side of Eq.~(\ref{eq:XE_deriv_ab_init})
remains valid when 
approximate functionals are used as long as the (now approximate) ensemble
energy $E^w$ is calculated variationally (i.e. by minimization over
densities) according to Eq.~(\ref{eq:Ew_KS-eDFT_ab_init}). This is due
to the stationarity of the (now approximate) minimizing
ensemble density $n^w$. As pointed
out previously,
in this case, $\ddroit E^w/\ddroit w$ might become $w$-dependent and
therefore,
for a given value of $w$, it may deviate from the slope obtained by
linear interpolation (right-hand side of Eq.~(\ref{eq:LIM_exact})). 

\subsection{Ensemble DFT for the two-site Hubbard model}

In the two-site Hubbard model~\cite{Hubbard_1}, the {\it ab initio} Hamiltonian
of Eq.~(\ref{eq:Hamilto_v}) is simplified as follows,
 \begin{eqnarray} 
 \hat{T}&\;\;\rightarrow\;\;& \hat{\mathcal{T}}=-t \sum_{\sigma=\uparrow,\downarrow}\left(\hat{a}_{0\sigma}^\dagger \hat{a}_{1\sigma} + \hat{a}_{1\sigma}^\dagger 
 \hat{a}_{0\sigma}\right)
,
\nonumber\\
\hat{W}_{\rm ee}&\;\;\rightarrow\;\;& 
\hat{U}=U\sum^1_{i=0}\hat{a}_{i\uparrow}^\dagger
\hat{a}_{i\uparrow}\hat{a}_{i\downarrow}^\dagger \hat{a}_{i\downarrow}  
,
\nonumber\\
\int \ddroit{\bf r}\;v
({\bf r})\hat{n}({\bf r})&\rightarrow&
\dfrac{\Delta v}{2}\left( \hat{n}_1 - \hat{n}_0 \right)
,
 \end{eqnarray} 
where operators are written in second quantization and the labels 0 and
1 refer to the first and second atomic site, respectively. 
The density operator on site $i$ reads $\hat{n}_i=
\sum_{\sigma=\uparrow,\downarrow}\hat{a}_{i\sigma}^\dagger\hat{a}_{i\sigma}$.
As shown in Refs.~\cite{deur2017exact} and~\cite{carrascal2015hubbard},
various correlation and density regimes can be explored by varying the
three parameters of the model, namely $t$ (the hopping parameter), $U$
(the strength of the on-site two-electron repulsion) and the local potential
parameter $\Delta v$ which controls the asymmetry of the model.
Following Ref.~\cite{deur2017exact}, we will describe in the rest of
this work a {\it two-electron} ensemble consisting of the ground- and
first-excited {\it singlet} states of the Hubbard dimer. The exact
energies $E_i$ ($i=0,1$), which are functions of $t$, $U$ and $\Delta v$, can be
determined analytically by solving the following third-order polynomial
equation~\cite{deur2017exact,carrascal2015hubbard,PRB16_Burke_thermal_DFT_dimer},    
\begin{eqnarray}\label{eq:3rdorder_energy_prb}
%\hspace{-0.8cm}
-4 t^{2} U + \left(4 t^{2}- U^{2}+\Delta v ^{2}\right)E_i+ 2 U E_i^{2} 
=E_i^{3}.
\end{eqnarray}
In this context, a trial density consists in principle of two
numbers, $n_0$ and $n_1$, which are the occupations of site 0 and 1,
respectively. In the particular case of two electrons, the density
can be reduced to a single occupation number $n=n_0$ since $n_1=2-n_0$. Consequently, for
a given external local potential $\Delta v=\Delta v_{\rm ext}$, the
exact ensemble energy 
$E^w=E^w\left(\Delta v_{\rm ext}\right)$ of the two-electron Hubbard dimer can be
expressed as follows in KS-eDFT, 
\begin{eqnarray}\label{eq:Ew_GOKDFT_HD}
E^{w} = \underset{n}{\rm inf}
\Big\{E^w_{\Delta v_{\rm ext}}(n)\Big\},
\end{eqnarray}
where the density-functional ensemble energy to be minimized reads%~\cite{deur2017exact}
\begin{eqnarray}\label{eq:densfun_Ew_GOKDFT}
E^w_{\Delta v_{\rm ext}}(n)&=& 
T_{\rm s}^{w}(n) + 
E_{{\rm
H}}(n)
+E^{w}_{{\rm
x}}(n)
+
E^{w}_{{\rm
c}}(n)
\nonumber\\
&&+ \Delta v_{\rm ext}\times(1-n)
,
\end{eqnarray}
in analogy with the {\it ab initio} expression of
Eq.~(\ref{eq:Ew_KS-eDFT_ab_init}).
Note that the $t$ and $U$ dependencies of the 
various density-functional energy contributions have been dropped for
clarity. Note also that the latter functionals are in fact {\it functions} of the occupation number $n$ that
will be referred to as density in the rest of this work. As shown in
Ref.~\cite{deur2017exact}, exact analytical expressions can be derived
for all functionals except the correlation one: 
\be
\label{eq:noninteractingenergyhubbarddimer}
T_{\rm s}^w(n) &=& -2t \sqrt{(1 - w)^2 - (1 - n)^2},
\\
E_{\rm H}(n)
&=&U\Big(1+(1-n)^2\Big),
\\
\label{eq:EEXX_fun_HD}
E^{w}_{\rm x}(n)&=&\dfrac{U}{2}\left[1+w-\dfrac{(3w-1)(1-n)^2}{(1-w)^2}\right]
\nonumber
\\
&&-E_{\rm H}(n).
\ee
As readily seen from Eq.~(\ref{eq:noninteractingenergyhubbarddimer}), a
density $n$ is ensemble non-interacting $v$-representable if 
\be\label{eq:non_int_vrep_cond}
\vert 1-n\vert \leq 1-w.
\ee
For densities in the latter range, the exact ensemble correlation energy can be obtained 
numerically as
follows~\cite{deur2017exact},
\be
E^{w}_{{\rm
c}}(n)
&=&
F^w(n)-T_{\rm s}^{w}(n)- 
E_{{\rm
H}}(n)
-E^{w}_{{\rm
x}}(n),
\ee
where, in analogy with the {\it ab initio} expression in Eq.~(\ref{universalfunctional}),  
\begin{eqnarray}
\label{universalfunctionalHD}
F^{w}(n) = \underset{\Delta v}{\rm sup}\Big\{ E^{w}(\Delta v) + \Delta v
\times(n - 1)\Big\}.
\end{eqnarray}
Note that, for a trial potential $\Delta v$, the ensemble energy
$E^{w}(\Delta v)$ is determined from Eq.~(\ref{eq:3rdorder_energy_prb}).
Obviously, for practical calculations, analytical DFAs
are preferable to numerical ones. Moreover,
developing a general strategy for the derivation of weight-dependent correlation
functionals that might also be applicable to {\it ab initio}
Hamiltonians is highly desirable. We will show
in the following how explicit correlation density functionals can be constructed by
expanding the Legendre--Fenchel transform of Eq.~(\ref{universalfunctionalHD}) 
in the vicinity of various density and correlation regimes. 

\section{Taylor expansions of the exact ensemble correlation functional}
\label{sec:Taylor_exp_various_regimes}
%\subsection{Weight-dependent density-functional approximations}

\subsection{Expansion around the symmetric 
case}\label{subsec:exp_in_delta}

For convenience we introduce the on-site repulsion $u=U/(2t)$, local
potential $\nu=\Delta v/(2t)$ and ensemble Legendre--Fenchel transform
\be
f^w(\delta)=F^w(1+\delta)/(2t)
\ee 
per unit of $2t$, thus leading to (see Eq.~(\ref{universalfunctionalHD}))
\begin{eqnarray}\label{eq:LF_over_2t}
{f}^w(\delta)&=&\underset{\mathbf{{\nu}}}{\rm sup} \Big\{
(1-w){e}_0({\nu})+w{e}_1({\nu})+ {\nu}\delta \Big\}
\nonumber\\
&=&
(1-w){e}_0\Big({\nu}(\delta)\Big)+w{e}_1\Big({\nu}(\delta)\Big)+
{\nu}(\delta)\delta
,
%\nonumber\\
\end{eqnarray}
where, for given values of $u$ and $\nu$, the individual energies
$e_i=E_i/(2t)\equiv e_i(\nu,u)$ of the ground- ($i=0$) and first-excited ($i=1$) singlet
states are, according to Eq.~(\ref{eq:3rdorder_energy_prb}), solutions of    
\begin{eqnarray}\label{eq:3rd_order_poly_eq_ener}
-u+e_i(1-u^2+\nu^2)+2ue_i^2=e_i^3.
\end{eqnarray}
Note that the $u$-dependence of $e_0$ and $e_1$ has been dropped in Eq.~(\ref{eq:LF_over_2t}) for
clarity. 
In order to expand the ensemble Legendre--Fenchel transform $F^w(n)$ around the
symmetric $n=1$ case, which is equivalent to expanding ${f}^w(\delta)$ around
$\delta=0$, 
\be
f^w(\delta)&=&f^w(0)+\delta\left.\dfrac{\ddroit f^w(\delta)}{\ddroit
\delta}\right|_{\delta=0}+\dfrac{\delta^2}{2}\left.\dfrac{\ddroit^2 f^w(\delta)}{\ddroit
\delta^2}\right|_{\delta=0}
\nonumber\\
&&+\mathcal{O}(\delta^3),
\ee
we need to calculate energy derivatives. Indeed, by using the
stationarity of the
maximizing potential ${\nu}(\delta)$ in Eq.~(\ref{eq:LF_over_2t}), we obtain  
\be\label{eq:1st_deriv_fw}
%\left.
\dfrac{\ddroit f^w(\delta)}{\ddroit
\delta}%\right|_{\delta=0}
={\nu}(\delta),
\ee
thus leading to
\be\label{eq:d2fwd2delta0}
\left.\dfrac{\ddroit^2 f^w(\delta)}{\ddroit
\delta^2}\right|_{\delta=0}=\left.\dfrac{\ddroit {\nu}(\delta)}{\ddroit
\delta}\right|_{\delta=0}.
\ee
The latter response of the potential (to deviations in  
density from the symmetric case) is determined from the stationarity
condition, which holds for {\it any} $\delta$,
\be\label{eq:stationarity_cond_LF}
(1-w)\left.\dfrac{ \partial e_0(\nu)}{\partial \nu}\right|_{\nu={\nu}(\delta)}
+w\left.\dfrac{ \partial e_1(\nu)}{\partial
\nu}\right|_{\nu={\nu}(\delta)}=-\delta,
\ee
thus giving after differentiation with respect to $\delta$,
\be\label{eq:lr_vector_pot}
\left.\dfrac{\ddroit {\nu}(\delta)}{\ddroit
\delta}\right|_{\delta=0}=-\left[(1-w)
\dfrac{ \partial^2
e_0(\nu)}{\partial \nu^2}
+
w\dfrac{ \partial^2
e_1(\nu)}{\partial \nu^2}
\right]_{\nu={\nu}(0)}^{-1}
. \nonumber \\
\ee
Differentiating Eq.~(\ref{eq:3rd_order_poly_eq_ener}) with respect to
$\nu$ gives
\be\label{eq:diff_nu_3rd_order_eq}
\dfrac{ \partial e_i(\nu)}{\partial
\nu}\times\Big[1-u^2+\nu^2+4ue_i(\nu)-3e_i^2(\nu)\Big]=-2\nu e_i(\nu)
,
\nonumber\\
\ee
which, when combined with Eqs.~(\ref{eq:1st_deriv_fw}) and (\ref{eq:stationarity_cond_LF}), leads to
the expected solution~\cite{deur2017exact},  
\be\label{eq:dfwddelta=0}
{\nu}(0)=0=\left.\dfrac{\ddroit f^w(\delta)}{\ddroit
\delta}\right|_{\delta=0}.
\ee
Similarly, by differentiating Eq.~(\ref{eq:diff_nu_3rd_order_eq}) with
respect to $\nu$ and using Eq.~(\ref{eq:lr_vector_pot}), we obtain (see Appendix~\ref{appendix:simplify_d2fwd2delta})
\be\label{eq:final_exp_d2fwd2w}
\left.\dfrac{\ddroit^2 f^w(\delta)}{\ddroit
\delta^2}\right|_{\delta=0}
&=&
\dfrac{g(u)}{2\Big(1+w\big[ug(u)-1\big]\Big)}
%\dfrac{ue-u^2-2}{2\Big(e-w\big[e+u(2+u^2-ue)\big]\Big)}
,\ee
where
\be\label{eq:def_g(u)}
g(u)=\left(\frac{u}{2}+\sqrt{1+\left(\frac{u}{2}\right)^2}\right)
\left[1+\left(\frac{u}{2}+\sqrt{1+\left(\frac{u}{2}\right)^2}\right)^2\right]
.
\nonumber\\
\ee
\iffalse%%%%
\be
e=e_0(0)=\dfrac{1}{2}\left(u-\sqrt{u^2+4}\right).
\ee 
\fi%%%%%%%%%%
Turning to the ensemble correlation energy (per unit of $2t$),
\be
e^w_{\rm
c}(\delta)=f^w(\delta)-f^w(\delta,u=0)-e^w_{\rm Hx}(\delta),
\ee 
where, according to Eq.~(\ref{eq:EEXX_fun_HD}),
\be\label{eq:ensHx_over_2t}
e^w_{\rm Hx}(\delta)=\dfrac{u}{2}\left[1 + w - \dfrac{(3w - 1)\delta^2}{(1 - w)^2} \right]
,
\ee 
we finally obtain from Eqs.~(\ref{eq:dfwddelta=0}) and (\ref{eq:final_exp_d2fwd2w}) the following expansion  
through second order
in $\delta$,
\be\label{eq:Taylor_exp_2nd_order_delta}
e^w_{\rm c}(\delta)&=&(1-w)\left[1-\sqrt{1+\left(\frac{u}{2}\right)^2}\right]
\nonumber\\
&&+\dfrac{\delta^2}{4}\left[\dfrac{g(u)}{1+w\big[ug(u)-1\big]}
-\dfrac{2\big(u+1-w(3u+1)\big)}{(1-w)^2}
\right]
\nonumber\\
&&+\mathcal{O}\left(\delta^4\right),
\ee     
where we used the simplified expression 
\be
f^w(0)=\dfrac{u(1+w)}{2}-(1-w)\sqrt{1+\left(\frac{u}{2}\right)^2}
,
\ee
which is deduced from Eqs.~(\ref{eq:LF_over_2t}),
(\ref{eq:dfwddelta=0}), (\ref{eq:e0_exp}), and (\ref{eq:e1_exp}).
Note that, as expected, the expansion 
obtained by Carrascal {\it et al.} around $n=1$ for the ground-state functional (see Eq.~(B.13) in
Ref.~\cite{carrascal2015hubbard} where $\rho$ corresponds to our
$\vert\delta\vert$) is recovered from
Eq.~(\ref{eq:Taylor_exp_2nd_order_delta}) when $w=0$.\\

Let us finally focus on the behavior of the expansion in
Eq.~(\ref{eq:Taylor_exp_2nd_order_delta}) when $\vert\delta\vert=1/2$
and $w=\frac{1}{2}-\eta$ where $\eta \ll 1$ (i.e. close to the equi-ensemble
case and far from the symmetric case). Truncation through second order in $\delta$ and first order in
$\eta$ gives
\be\label{eq:deltaPT2_0.5_minus_eta}
&&2e^{w=\frac{1}{2}-\eta}_{\rm c}\left(\pm\frac{1}{2}\right)
\approx
\nonumber\\
%\dfrac{1}{2}
&&\Bigg[\dfrac{1}{2}-\dfrac{1}{\dfrac{u}{2}+\sqrt{1+\left(\dfrac{u}{2}\right)^2}}
%\nonumber\\
%&&
+\dfrac{g(u)}{4\left(1+ug(u)\right)}
\Bigg]
\nonumber\\
&&+
%\dfrac
{\eta}
%{2}
\left[
3-5u-2\sqrt{1+\left(\dfrac{u}{2}\right)^2}
-\dfrac{g(u)\left(1-ug(u)\right)}{2\left(1+ug(u)\right)^2}\right]
.\ee
Interestingly, the latter correlation energy expression will vary as
follows in the strongly correlated limit,
\be\label{eq:deltaPT2_0.5_minus_eta_u_infty}
e^{w=\frac{1}{2}-\eta}_{\rm c}\left(\pm\frac{1}{2}\right)\underset{u\rightarrow+\infty}{\approx}
\dfrac{1}{4}-3u\eta,
\ee 
and, as readily seen, an unphysical positive result is obtained when
$\eta=0$. In other words, the expansion in
Eq.~(\ref{eq:Taylor_exp_2nd_order_delta}) is expected to fail 
in practice if calculations are performed with $w=1/2$ in such regimes
of density and correlation. Note also that, when $\eta>0$, the expansion
in Eq.~(\ref{eq:deltaPT2_0.5_minus_eta_u_infty}) becomes
\be\label{eq:deltaPT2_0.5_minus_eta_posi_u_infty}
e^{w=\frac{1}{2}-\eta}_{\rm c}\left(\pm\frac{1}{2}\right)\underset{u\rightarrow+\infty}{\approx}
-3u\eta,
\ee 
which is actually incorrect, as will be discussed further in
Sec.~\ref{subsec:strong_corr_regime}.
%%%%%%%%%%%%%%%%%%%%%%%%%%%%%%%
\iffalse%%%
Finally, using the expansion in also leads to      
\be
-\left.\dfrac{\partial^2 e^{w=\frac{1}{2}-\eta}_{\rm
c}\left(\delta\right)}{\partial\delta\partial
\eta}\right|_{\eta=0^+,\vert\delta\vert=1/2}\underset{u\rightarrow+\infty}{\approx}
\ee
\fi%%%%
\iffalse%%%%%%%%%%%%%%%%%%
\be
e_{\rm c}(n)&=&\left[1-\sqrt{1+\left(\frac{u}{2}\right)^2}\right]
\nonumber\\
&&+\delta^2\left[-\dfrac{1}{2}+\left(\frac{u}{2}\right)^3\right]
\ee
\fi%%%%

\subsection{Expansion in the weakly correlated regime}

By following the same strategy as in Sec.~\ref{subsec:exp_in_delta}, we will expand in this
section
the ensemble Legendre--Fenchel transform around $u=0$ for a {\it fixed} ensemble
non-interacting $v$-representable deviation $\delta$
from the symmetric case, i.e. any deviation such that (see Eq.~(\ref{eq:non_int_vrep_cond}))
\be\label{eq:non_int_v_represent_cond}
\vert\delta\vert\leq
1-w.
\ee
For clarity, we will make both $u$- and
$\delta$-dependencies explicit in Eq.~(\ref{eq:LF_over_2t}), thus leading to    
\begin{eqnarray}\label{eq:LF_over_2t_u_dependence}
{f}^w(\delta,u)&=&\underset{\mathbf{{\nu}}}{\rm sup} \Big\{
(1-w){e}_0({\nu},u)+w{e}_1({\nu},u)+ {\nu}\delta \Big\}
\nonumber\\
&=&
(1-w){e}_0\Big({\nu}^w(\delta,u),u\Big)+w{e}_1\Big({\nu}^w(\delta,u),u\Big)
\nonumber\\
&&+
{\nu}^w(\delta,u)\delta
,
%\nonumber\\
\end{eqnarray}
and the Taylor expansion
\be
f^w(\delta,u)&=&f^w(\delta,0)+u\left.\dfrac{\ddroit f^w(\delta,u)}{\ddroit
u}\right|_{u=0}+\dfrac{u^2}{2}\left.\dfrac{\ddroit^2 f^w(\delta,u)}{\ddroit
u^2}\right|_{u=0}
\nonumber\\
&&+\mathcal{O}(u^3),
\ee
where, according to Eq.~(\ref{eq:noninteractingenergyhubbarddimer}),
\be\label{eq:tsw_over_2t}
f^w(\delta,0)&=&T^w_{\rm s}(1+\delta)/(2t)
\nonumber\\
&=&-\sqrt{(1-w)^2-\delta^2}
,
\ee
with the corresponding maximizing (KS) potential~\cite{deur2017exact}
\be\label{eq:KS_pot_over_2t}
{\nu}^w(\delta,0)=\dfrac{\delta}{\sqrt{(1-w)^2-\delta^2}}
.
\ee
From the stationarity condition in Eq.~(\ref{eq:stationarity_cond_LF}),
which holds for {\it any} $u$ and that, for clarity, we will rewrite as
follows,
%\iffalse%%%%%%%%%%
\be\label{eq:stat_cond_weakly_corr}
\left[(1-w)\dfrac{\partial e_0({\nu},u)}{\partial \nu}
+w\dfrac{\partial e_1({\nu},u)}{\partial \nu}
\right]_{\nu={\nu}^w(\delta,u)}
=-\delta,
\ee
%\fi%%%%%%%%%%
it comes
\be\label{eq:dfw_over_du}
\dfrac{\ddroit f^w(\delta,u)}{\ddroit
u}=
\left[(1-w)\dfrac{\partial e_0({\nu},u)}{\partial u}
+w\dfrac{\partial e_1({\nu},u)}{\partial u}
\right]_{\nu={\nu}^w(\delta,u)}
,
\nonumber
\\
\ee
and
\be\label{eq:d2fw_du2_0}
&&\left.\dfrac{\ddroit^2 f^w(\delta,u)}{\ddroit
u^2}\right|_{u=0}
=
\Big[(1-w)\dfrac{\partial^2 e_0({\nu},u)}{\partial \nu\partial u}
\nonumber\\
&&+w\dfrac{\partial^2 e_1({\nu},u)}{\partial \nu\partial u}
\Big]_{\nu={\nu}^w(\delta,0),u=0}
%\nonumber\\
%&&
\times\left.\dfrac{\partial {\nu}^w(\delta,u)}{\partial u}\right|_{u=0}
\\
&&
+
\left[
(1-w)\dfrac{\partial^2 e_0({\nu},u)}{\partial u^2}
+w\dfrac{\partial^2 e_1({\nu},u)}{\partial u^2}
\right]_{\nu={\nu}^w(\delta,0),u=0}
, \nonumber
\ee
where the linear response of the potential
$
%\left.
{\partial {\nu}^w(\delta,u)}/{\partial u}
%\right|_{u=0}
$ is
determined by
differentiating Eq.~(\ref{eq:stat_cond_weakly_corr}) with respect to $u$,
thus leading to
\be\label{eq:LR_vec}
%\left.
\dfrac{\partial {\nu}^w(\delta,u)}{\partial u}
%\right|_{u=0}
=
-
\left.
\dfrac{(1-w)\dfrac{\partial^2 e_0({\nu},u)}{\partial\nu\partial u}
+w\dfrac{\partial^2 e_1({\nu},u)}{\partial \nu\partial
u}}{(1-w)\dfrac{\partial^2 e_0({\nu},u)}{\partial \nu^2}
+w\dfrac{\partial^2 e_1({\nu},u)}{\partial \nu^2}}
\right|_
%{\nu={\nu}(\delta,0),u=0}
{\nu={\nu}^w(\delta,u)}
.
\nonumber\\
\ee
As shown in Appendix~\ref{appendix:simp_deriv_u}, simple expressions
(in terms of $\delta$ and $w$) can be obtained for all 
energy derivatives, thus showing that the exact ensemble
Hx energy is recovered through first order in $u$, as expected, while the
ensemble correlation energy (obtained through second order in $u$) reads 
\be\label{eq:Taylor_u_second_order}
\dfrac{e^w_{\rm
c}(\delta,u)}{u^2}
&=&
\dfrac{1}{2}\left.\dfrac{\ddroit^2 f^w(\delta,u)}{\ddroit
u^2}\right|_{u=0}+
\mathcal{O}\left(u\right)
\nonumber\\
&=&-\dfrac{\left[(1-w)^2-\delta^2\right]^{3/2}}{8(1-w)^2}
\nonumber\\
&&\times\left[1+\dfrac{\delta^2}{(1-w)^2}\left(3-\dfrac{4(1-3w)^2}{(1-w)^2}\right)\right]
\nonumber
\\
&&
+\mathcal{O}\left(u\right)
.
\ee
Note that, by inserting the following expansion of $g(u)$ (see
Eq.~(\ref{eq:def_g(u)})) into Eq.~(\ref{eq:Taylor_exp_2nd_order_delta}),
\be
g(u)=2(1+u)+\dfrac{5}{4}u^2+\mathcal{O}\left(u^3\right),
\ee
or by expanding the expression in Eq.~(\ref{eq:Taylor_u_second_order})
through second order in $\delta$,
we recover the same
expression, as expected~\cite{carrascal2015hubbard},
\be\label{eq:ec_over_u2_weakly_corr}
\dfrac{e^w_{\rm
c}(\delta,u)}{u^2}
%\underset{\delta\rightarrow 0}{\sim}
&=&-\dfrac{(1-w)}{8}+\dfrac{\delta^2\left[8(1-3w)^2-3(1-w)^2\right]}{16(1-w)^3}
\nonumber\\
&&+\mathcal{O}\left(u,\delta^4\right).
\ee
As readily seen from Eq.~(\ref{eq:ec_over_u2_weakly_corr}), in this regime of correlation, the ensemble
density-functional correlation energy will be concave when
$\frac{21-4\sqrt{6}}{69}\approx0.16\leq w\leq 
\frac{21+4\sqrt{6}}{69}\approx 0.45$,
and convex otherwise.     

\iffalse%%%%%%%%%%%%%%%%

The key quantity to model (maybe not ...) is the deviation $\Delta e^w_{\rm
c}(n)=e^w_{\rm c}(n)-e_{\rm c}(n)$ of the ($w$-dependent)
ensemble correlation density-functional energy from the conventional
ground-state functional 
($w=0$). Following Carrascal {\it et al.}~\cite{carrascal2015hubbard},
we derived in the appendices Taylor expansions of $\Delta e^w_{\rm
c}(n)$ around $n=1$ (for all correlation regimes), thus leading to the final
expression,
\be
&&\Delta e^w_{\rm
c}(n)=\dfrac{w}{2}\left(\sqrt{u^2+4}-2\right)
\nonumber\\
&&
+w\delta^2\left[\chi^w(u)+\dfrac{\Big(u-1+w(u+1)\Big)}{2(1-w)^2}\right]
%\nonumber\\
%&&
+\mathcal{O}\left(\delta^4\right)
,
\ee
where $\delta =(n-1)$ and
\be
\chi^w(u)&=&\dfrac{-\alpha(u)\sqrt{u^2+4}}
{\left(\sqrt{u^2+4}-u\right)^2\left[4+w\alpha(u)\right]}
,
\nonumber\\
\alpha(u)&=&(u^2+2)\left(u+\sqrt{u^2+4}\right)^2-12.
\ee 
Similarly, the following expansions, 
\be
&&\dfrac{8}{u^2}\Delta e^w_{\rm
c}(n)
-\left(1-\delta^2\right)^{5/2}
\nonumber\\
=&&-\dfrac{\left[(1-w)^2-{\delta^2}\right]^{3/2}}{(1-w)^6}
\nonumber\\
&&\times\Big[(1-w)^4-\delta^2(33w^2-18w+1)\Big]
\nonumber\\
&&+\mathcal{O}(u)
\ee

as well as $u=0$ and
$u\rightarrow$ (for all density regimes), 
\fi%%%%%%%%%%%%%%

\subsection{Strongly correlated limit}\label{subsec:strong_corr_regime}

\begin{figure}
\resizebox{0.49\textwidth}{!}{
\includegraphics[scale=1]{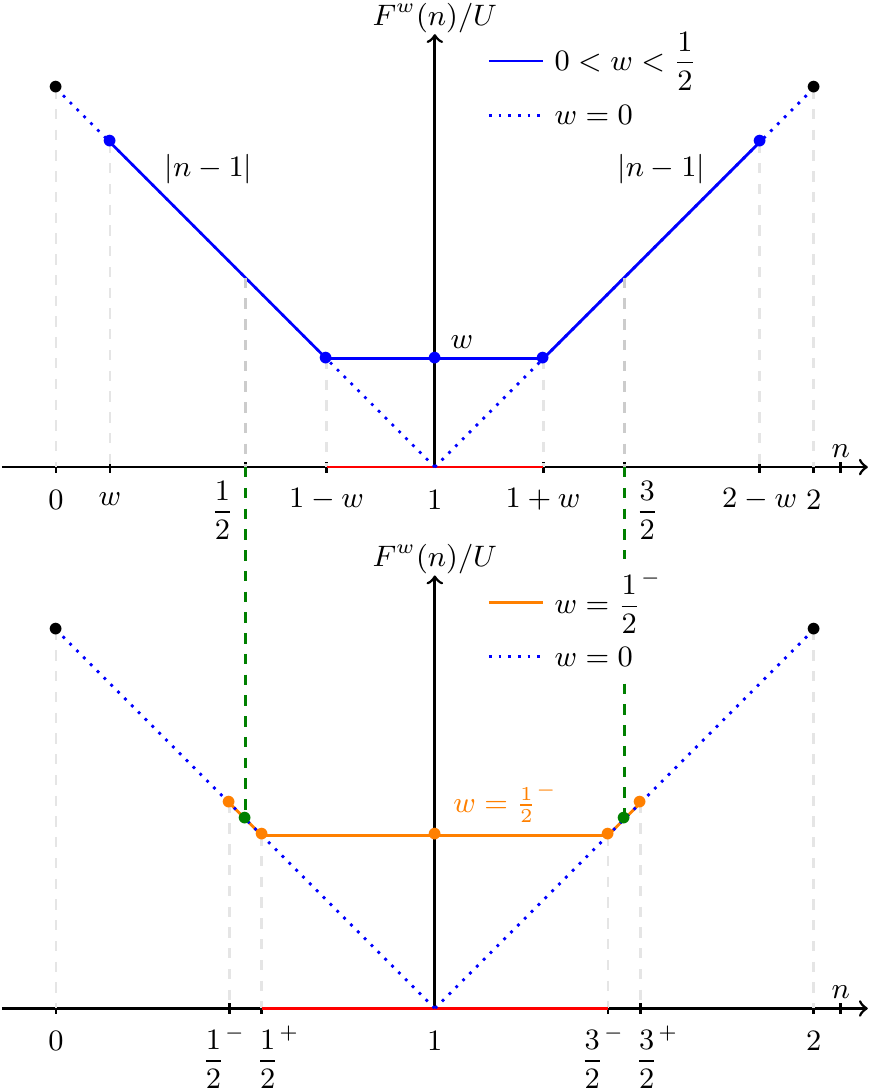}
}
\caption{Graphical representation of the exact ensemble functional $F^w(n)$ in the
$U/t\rightarrow+\infty$ limit. The general case where the ensemble weight
is in the range $0<w<1/2$ is shown in the top panel. The particular case
$w = (1/2)^-=(1/2)-\eta$ where $\eta \rightarrow 0^+$ is shown in the
bottom panel.
%For a density $n$, we denote $n^\pm=n\pm\eta$. 
The (weight-dependent)
density domain where the ensemble functional differs from the ground-state one is
shown in red.
}
\label{fig:tikz2}
\end{figure}

Let us, for convenience, consider the Legendre--Fenchel transform in
Eq.~(\ref{eq:LF_over_2t}) per unit of $u$,
\begin{eqnarray}\label{eq:LF_over_u}
\overline{f}^w(\delta)&=&{f}^w(\delta)/u
\nonumber\\
&=&\underset{\mathbf{\overline{\nu}}}{\rm sup} \Big\{
(1-w)\overline{e}_0(\overline{\nu})+w\overline{e}_1(\overline{\nu})+
\overline{\nu}\delta \Big\},
%\nonumber\\
%&=&(1-w)\overline{e}_0+w\overline{e}_1+ \overline{\nu}(n-1)
\end{eqnarray}
where $\overline{\nu}=\nu/u$ and, according to
Eq.~(\ref{eq:3rd_order_poly_eq_ener}), the $\overline{\nu}$-dependent ground-
and first-excited-state energies are, in the strongly correlated limit
($u\rightarrow+\infty$), solutions of
\begin{eqnarray}
\overline{e}_i\times(\overline{\nu}^2-1)+2\left(\overline{e}_i\right)^2=\left(\overline{e}_i\right)^3
,
\end{eqnarray}
thus leading to
\begin{eqnarray}
\overline{e}_0(\overline{\nu})&=&
{\rm inf}
\Big\{0,1-\left\vert\overline{\nu}\right\vert\Big\},
\nonumber\\
\overline{e}_1(\overline{\nu})&=&
{\rm sup}
\Big\{0,1-\left\vert\overline{\nu}\right\vert\Big\}.
\end{eqnarray}
Therefore,
\be
\overline{f}^w(\delta)\underset{u\rightarrow+\infty}{\longrightarrow}
\underset{
}{\rm sup} \Big\{
\overline{f}^{w,\leq}(\delta),\overline{f}^{w,\geq}(\delta)
\Big\},
\ee
where, according to Eq.~(\ref{eq:non_int_v_represent_cond}),  
\begin{eqnarray}
\overline{f}^{w,\geq}(\delta)&=&
\underset{\vert{\overline{\nu}\vert}\geq 1}{\rm sup} \Big\{
(1-w)\left(1-\left\vert\overline{\nu}\right\vert\right)+
\overline{\nu}\delta \Big\}
\nonumber
\\
&=&{\rm sup}\Big\{\delta,-\delta\Big\},
\end{eqnarray}
and 
\begin{eqnarray}
\overline{f}^{w,\leq}(\delta)&=&
\underset{\vert{\overline{\nu}\vert}\leq 1}{\rm sup} \Big\{
w\left(1-\left\vert\overline{\nu}\right\vert\right)+
\overline{\nu}\delta \Big\}
\nonumber
\\
&=&{\rm sup}\Big\{\delta,w,-\delta\Big\}.
\end{eqnarray}
We conclude that
\begin{eqnarray}\label{eq:fw_over_u_strongcorr}
\overline{f}^{w}(\delta)
&\underset{u\rightarrow+\infty}{\longrightarrow}&
{\rm sup}\Big\{\delta,w,-\delta\Big\}.
\end{eqnarray}
A graphical summary of Eq.~(\ref{eq:fw_over_u_strongcorr}) is
given in Fig.~\ref{fig:tikz2}. As readily seen, the
functional will return $w$ for densities in the range $\vert\delta\vert\leq w$,
thus leading to (see
Eqs.~(\ref{eq:ensHx_over_2t}) and~(\ref{eq:tsw_over_2t})), 
\be\label{eq:ec_exp_large_u_delta_inf_w}
\dfrac{e^w_{\rm
c}(\delta)}{u}\underset{u\rightarrow+\infty}{\longrightarrow}-\dfrac{1}{2}\left[(1-w)-\dfrac{(3w-1)\delta^2}{(1-w)^2}\right].
\ee
\\
Interestingly, the Taylor expansion of the
ensemble correlation energy through second order in $\delta$ (see
Eq.~(\ref{eq:Taylor_exp_2nd_order_delta})) is becoming exact in the
strongly correlated limit for densities in the range $\vert\delta\vert\leq w$. 
As readily seen from Eq.~(\ref{eq:ec_exp_large_u_delta_inf_w}), in this
regime of correlation, the density-functional ensemble correlation
energy is concave if $0< w\leq 1/3$ and convex otherwise. 
\iffalse%%%%%%%%%%
\manu{
Note that for $w=1/2$ and $\vert\delta\vert=\frac{1}{2}-\eta_1$
($\eta_1>0$) we obtain
\be
e^{w=1/2}_{\rm
c}\left(\frac{1}{2}-\eta_1\right)\underset{u\rightarrow+\infty}{\sim}\eta_1(\eta_1-1)u.
\ee
}
Note that, when
$\vert\delta\vert=w$,    
\be
\dfrac{e^w_{\rm
c}(\pm w)}{u}\underset{u\rightarrow+\infty}{\longrightarrow}
\dfrac{(2w-1)\left[(4w-1)^2+7\right]}{16(1-w)^2}
,
\ee
which becomes zero in the particular case $w=1/2$.\\
\fi%%%%%%%%%
From 
Eqs.~(\ref{eq:non_int_v_represent_cond}) and
(\ref{eq:fw_over_u_strongcorr}), we finally see that,
as expected~\cite{deur2017exact},
the ensemble functional equals the ground-state
one for densities in the range
$w\leq\vert\delta\vert\leq 1-w$:
\be
\overline{f}^{w}(\delta)=\overline{f}^{w=0}(\delta)=\vert\delta\vert
.\ee
As a
result, in this density regime, the ensemble correlation energy reads  
\be\label{eq:ec_exp_large_u_delta_sup_w}
\dfrac{e^w_{\rm
c}(\delta)}{u}\underset{u\rightarrow+\infty}{\longrightarrow}\vert\delta\vert-\dfrac{1}{2}\left[(1+w)-\dfrac{(3w-1)\delta^2}{(1-w)^2}\right].
\ee
In the particular case $\vert\delta\vert=1/2$ and $w=\frac{1}{2}-\eta$
($\eta>0$) which has been considered previously in
Sec.~\ref{subsec:exp_in_delta},
Eq.~(\ref{eq:ec_exp_large_u_delta_sup_w}) is applicable, thus leading to
the following Taylor expansion through first order in $\eta$,  
\be\label{eq:exp_u_infty_exact_delta_0.5}
{e^{w=\frac{1}{2}-\eta}_{\rm
c}\left(\pm\frac{1}{2}\right)}\underset{u\rightarrow+\infty}{\approx}
-2\eta u.
\ee
Note that, in contrast to the expression in
Eq.~(\ref{eq:deltaPT2_0.5_minus_eta_u_infty}), the exact expression in
Eq.~(\ref{eq:exp_u_infty_exact_delta_0.5}) always gives a negative
correlation energy, as it should. 
%Interestingly, the correlation energy
%of the equiensemble
%($\eta=0$) equals zero in this regime of density and correlation.
Note
also that the expansion in
Eq.~(\ref{eq:deltaPT2_0.5_minus_eta_posi_u_infty}) is indeed
incorrect.\\   

For a {\it fixed} density deviation $\delta$, the ensemble
correlation energy $e^w_{\rm
c}(\delta)$ becomes a function of $w$ whose domain of definition
is given by the non-interacting ensemble representability condition,
i.e.
$0\leq w\leq 1-\vert\delta\vert$. In the particular case
\be
1-\vert\delta\vert\leq\vert\delta\vert\leq 1-w,
\ee
or, equivalently,
\be
1/2\leq \vert\delta\vert\leq1-w
,\ee the expression in
Eq.~(\ref{eq:ec_exp_large_u_delta_sup_w}) applies and, consequently,
\be\label{eq:DD_outside}
\dfrac{1}{u}\dfrac{\partial e^w_{\rm
c}(\delta)}{\partial w}
\underset{u\rightarrow+\infty}{\longrightarrow}
-\dfrac{1}{2}\left[1-\dfrac{\delta^2(1+3 w)}{(1- w)^3}\right].
\ee
On the other hand, if $\vert\delta\vert\leq 1-\vert\delta\vert$ or,
equivalently,
\be\label{eq:cond_va_delta_lower1/2}
\vert\delta\vert\leq 1/2
,\ee then two
cases must be distinguished. Either $0\leq w\leq \vert\delta\vert$
and, in this case, Eq.~(\ref{eq:DD_outside}) applies, or
$\vert\delta\vert\leq w\leq1-\vert\delta\vert$ and then Eq.~(\ref{eq:ec_exp_large_u_delta_inf_w})
applies, thus leading to
\be\label{eq:DD_inside}
\dfrac{1}{u}\dfrac{\partial e^w_{\rm
c}(\delta)}{\partial w}
\underset{u\rightarrow+\infty}{\longrightarrow}
\dfrac{1}{2}\left[1+\dfrac{\delta^2(1+3 w)}{(1-w)^3}\right].
\ee
Note that, as readily seen from Eqs.~(\ref{eq:DD_outside}) and
(\ref{eq:DD_inside}), and expected from Ref.~\cite{deur2017exact}, for densities
that fulfill the condition in Eq.~(\ref{eq:cond_va_delta_lower1/2}), there is a jump in the ensemble correlation energy
derivative with respect to the weight $w$ when the latter crosses $\vert\delta\vert$:  
\be
\left[
\left.\dfrac{\partial e^w_{\rm
c}(\delta)}{\partial w}
\right|_{w=\vert\delta\vert^+}
-\left.\dfrac{\partial e^w_{\rm
c}(\delta)}{\partial w}
\right|_{w=\vert\delta\vert^-}
\right]
\underset{u\rightarrow+\infty}{\longrightarrow}
u.
\ee
Let us finally consider the particular case of the equi-ensemble
($w=1/2$) for which the derivative of the ensemble
correlation energy must be taken at $w=\frac{1}{2}-\eta$ where
$\eta\rightarrow0^+$. 
In the strongly correlated limit (which also corresponds to the atomic
$t=0$ limit) we should have
\be\label{eq:DD_0.5_minus_u_infty}
\dfrac{1}{u}\left.\dfrac{\partial e^w_{\rm
c}\left(\delta=\pm\frac{1}{2}\right)}{\partial w}
\right|_{w=\frac{1}{2}^-}
%=
\underset{u\rightarrow+\infty}{\longrightarrow}
2,
\ee
while the expression in Eq.~(\ref{eq:DD_inside}) predicts the
(unphysical) result 
\be\label{eq:DD_0.5_plus_u_infty}
\dfrac{1}{u}\left.\dfrac{\partial e^w_{\rm
c}\left(\delta=\pm\frac{1}{2}\right)}{\partial w}
\right|_{w=\frac{1}{2}^+}
%=
\underset{u\rightarrow+\infty}{\longrightarrow}
3.
\ee
Note that Eqs.~(\ref{eq:DD_0.5_minus_u_infty}) and
(\ref{eq:DD_0.5_plus_u_infty}) are in agreement with
Eqs.~(\ref{eq:deltaPT2_0.5_minus_eta_posi_u_infty}) and
(\ref{eq:exp_u_infty_exact_delta_0.5}).

\iffalse%%%%%%%%% discussion about the DD %%%%%
From the previous discussion we can see that, if $\frac{1}{2}-\eta\leq\vert\delta\vert\leq 1/2$, 
Eq.~(\ref{eq:DD_outside}) can be applied, thus leading to 
%, which is the case when
%$\vert\delta\vert=1/2$, 
\be\label{eq:DD_outside_w1/2}
\dfrac{1}{u}\left.\dfrac{\partial e^\xi_{\rm
c}(\delta)}{\partial\xi}
\right|_{\delta=\frac{1}{2}^-}
\underset{\substack{u\rightarrow+\infty\\\xi\rightarrow\frac{1}{2}^-}}{\longrightarrow}
2.
\ee
Note that, by applying Eq.~(\ref{eq:DD_inside}) instead, 
we overestimate the ensemble correlation energy derivative 
by 50\%, which gives a substantial error when $u$ increases. The latter
equation is indeed applicable only when 
$\vert\delta\vert\leq\frac{1}{2}-\eta$.

A better (?) explaination follows:

\be
\dfrac{1}{u}\left.\dfrac{\partial e^\xi_{\rm
c}\left(\frac{1}{2}-\eta\right)}{\partial\xi}
\right|_{\xi=(\frac{1}{2}-\eta)^+}
\underset{\substack{u\rightarrow+\infty\\\eta\rightarrow0^+}}{\longrightarrow}3
\ee

\be
\dfrac{1}{u}\left.\dfrac{\partial e^\xi_{\rm
c}\left(\frac{1}{2}-\eta\right)}{\partial\xi}
\right|_{\xi=(\frac{1}{2}-\eta)^-}
\underset{\substack{u\rightarrow+\infty\\\eta\rightarrow0^+}}{\longrightarrow}2
\ee

\be
\dfrac{1}{u}\left.\dfrac{\partial e^\xi_{\rm
c}\left(\frac{1}{2}-\eta_1\right)}{\partial\xi}
\right|_{\xi=(\frac{1}{2}-\eta)^-}
\underset{\substack{u\rightarrow+\infty\\\eta\rightarrow0^+}}{\longrightarrow}2
\ee

\fi%%%%%%%%%%%%%

\section{Density-functional approximations and computational details}
\label{sec:comput_details}

A summary of the various DFAs that will be
tested on the Hubbard dimer in Sec.~\ref{sec:Results} is given here. The simplest approximation consists in
using the (weight-independent) ground-state (GS) xc functional, 
\be
E^w_{\rm
x}(n)\rightarrow E^{w=0}_{\rm
x}(n)
\hspace{0.2cm} \mbox{and} \hspace{0.2cm}
E^w_{\rm
c}(n)\rightarrow E^{w=0}_{\rm
c}(n).
\ee
It will be referred to as GSxc. The other approximations will all use the
(weight-dependent) ensemble exact exchange functional (see
Eq.~(\ref{eq:EEXX_fun_HD})). The ensemble exchange-only approximation
($E^w_{\rm
c}(n)\rightarrow 0$) will be referred to as EEXX. The ensemble correlation
energy will then be modeled either at the (weight-independent) ground-state
level,    
\be
E^w_{\rm
c}(n)\rightarrow 
E^{w=0}_{\rm
c}(n),
\ee
thus giving the GSc approximation, or with weight dependent functionals.
In the latter case, we will use the perturbation theory
expansion through second order (PT2) in the density deviation
$\delta=n-1$ from the symmetric case [the expansion is given in
Eq.~(\ref{eq:Taylor_exp_2nd_order_delta}) and will be referred to as
$\delta$-PT2] as well as the PT2 expansion 
in the weakly correlated regime, i.e. around $u=U/(2t)=0$ 
[the expansion is given in Eq.~(\ref{eq:Taylor_u_second_order}) and
will be referred to as $u$-PT2].
All calculations have been performed with $2t=1$. The accurate parameterization
of Carrascal~{\it et al.} (see Eqs.~(102)-(115) in
Refs.~\cite{carrascal2015hubbard,carrascal2016corrigendum}) has been used for the ground-state correlation
functional in GSxc and GSc calculations. Excitation energies have been
computed within the various approximations either by differentiation
(see Eq.~(\ref{eq:XE_deriv_ab_init})) or by linear interpolation (see
Eq.~(\ref{eq:LIM_exact})). In the former case, the excitation energy
reads as follows, according to Eq.~(\ref{eq:noninteractingenergyhubbarddimer}), 
\be\label{eq:dEw_dw_HD}
\dfrac{\ddroit E^w}{\ddroit w}&=&
\dfrac{2t(1-w)}{\sqrt{(1-w)^2-(1-n^w)^2}}
+\left.
\dfrac{\partial E_{\rm x}^w(n)}{\partial
w}
\right|_{n=n^w}
\nonumber\\
&&+
\left.\dfrac{\partial E_{\rm c}^w(n)}{\partial
w}
\right|_{n=n^w}
.
\ee

%The analytical expressions of the xc-DD for $\delta$-PT2 and $u$-PT2
%are provided in the appendix (Sec.~\ref{sec:appendix:de_dw})
%and are used to compute the excitation energies as described in
%Eq.~(\ref{eq:omega=derivative_w}).
\section{Results and discussion}\label{sec:Results}

In practical DFT calculations, the error in the energy is 
not only due to the approximate functional that is employed. It also depends on the
deviation from the exact result of the density
obtained by the minimization in Eq.~(\ref{eq:Ew_GOKDFT_HD}),
which is formally equivalent to solving the ensemble KS equations
self-consistently. Therefore, in the following, we will distinguish the so-called
functional driven error (Sec.~\ref{subsec:testing_DFA}), which is 
evaluated for a {\it fixed} density $n$, from the density driven one,
which will be discussed in the rest of this section. 

\subsection{
DFAs and functional driven error
%Testing the density-functional approximations
}\label{subsec:testing_DFA}

\begin{figure*}
\resizebox{1\textwidth}{!}{
\includegraphics[scale=1]{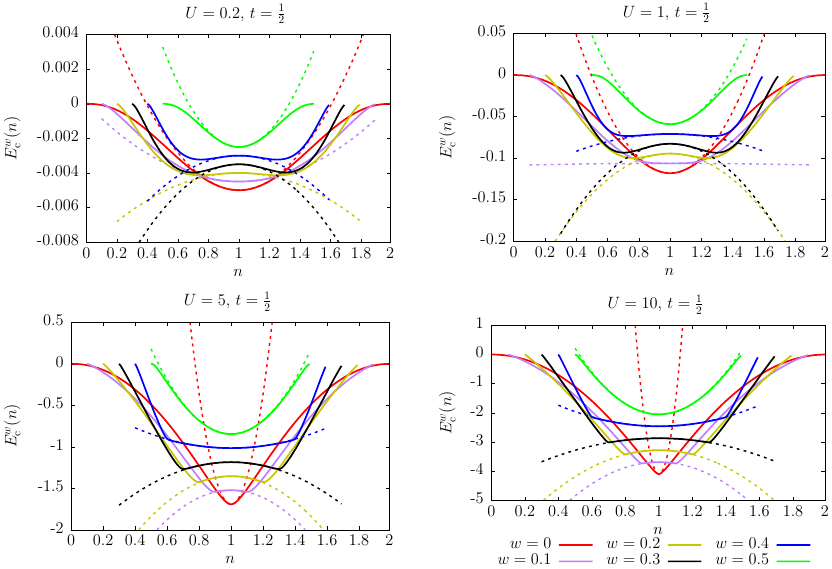}
}
\caption{$\delta$-PT2 correlation energy (dashed lines) plotted as a function of the density
for various correlation regimes and ensemble weights. Comparison is made with the exact
results (solid lines) of Ref.~\cite{deur2017exact}.
}
\label{fig:EcdeltaPT2}
\end{figure*}

\begin{figure}
\resizebox{0.49\textwidth}{!}{
\includegraphics[scale=1]{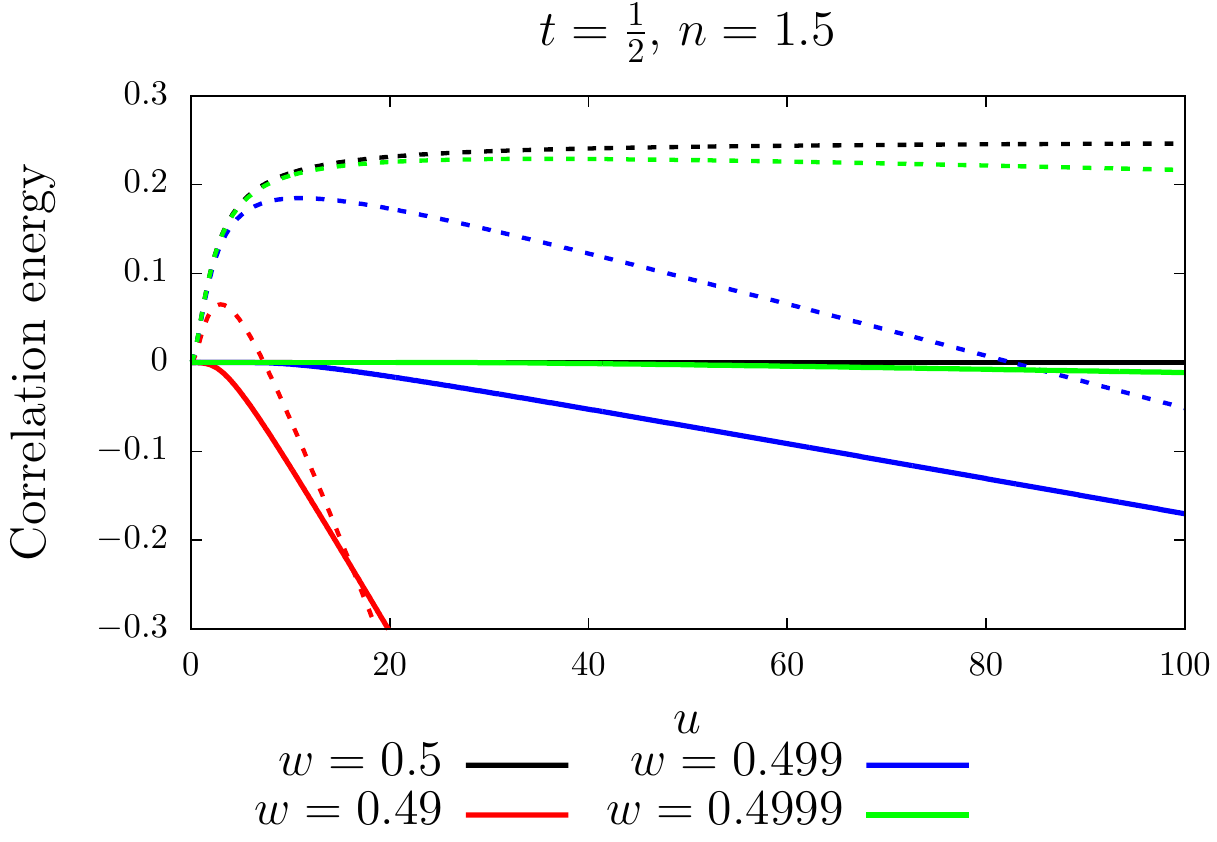}
}
\caption{$\delta$-PT2 correlation energy (dashed lines) plotted as a function of $u=U/(2t)$ for $n =
1.5$ and weights close to 1/2. Comparison is made with the exact
results (solid lines).}
\label{fig:positivecorrelationn15}
\end{figure}

Functional driven errors have already been studied in Ref.~\cite{deur2017exact} for both
GSc and GSxc approximations. Density functional correlation energies
obtained at the $\delta$-PT2 level (see Eq.~\ref{eq:Taylor_exp_2nd_order_delta}) are shown in Fig.~\ref{fig:EcdeltaPT2}.  
As expected, accurate correlation energies are obtained around $n=1$. We
also observe the changes in convexity when $w$ increases in both weakly
and strongly correlation regimes, as predicted by
Eqs.~(\ref{eq:ec_over_u2_weakly_corr}) and
(\ref{eq:ec_exp_large_u_delta_inf_w}), respectively. A major drawback of
the $\delta$-PT2 approximation is that it gives a non-zero (even
positive) correlation energy at the border of the $v$-representability 
domain, which is of course unphysical (see Appendix~\ref{appendix:Ec_border}). 
%(see Sec.~\ref{subsec:exp_in_delta}). 
In the light of Fig.~\ref{fig:EcdeltaPT2}, it is clear that $\delta$-PT2 should only be applied to
equi-ensembles (i.e. for $w=1/2$). Even though, in that case, accurate correlation energies are obtained
for a larger range of densities, in particular in the strongly
correlated regime, a spurious positive contribution remains when $n=1/2$
or $n=3/2$ as $U$
increases, as expected from
Eq.~(\ref{eq:deltaPT2_0.5_minus_eta_u_infty}) and illustrated in
Fig.~\ref{fig:positivecorrelationn15}.\\  
\iffalse%%%%%%%
 More precisly the quadratic therm is studied.
Writting an expansion for $u \rightarrow 0$ in the Eq. (\ref{eq:Taylor_exp_2nd_order_delta})
the following expression is obtained:
\begin{eqnarray}
 -\frac{1}{16}\frac{69w^{2} - 42w + 5}{(w-1)^{3}}u^{2} + \mathcal{O}\left(u^{3}\right)
\end{eqnarray}
and for $u \rightarrow +\infty$:
\begin{eqnarray}
\frac{1}{2(w-1)} + \frac{1}{2} \frac{3w-1}{(w-1)^{2}}u + \frac{1}{4wu} + \mathcal{O}\left(\frac{1}{u^3}\right).
\end{eqnarray}
These two expansions are vanished for exactly two values of $w$, this is why two changes 
of curvature are observed in the Fig. (\ref{fig:EcdeltaPT2}) when $w$ increases.
In particular if $w = 0.5$, in the weakly correlated regime the quadratic term is equal to
$2.5u^2$ and in the strongly correlated regime it is $u + \frac{1}{2u} -1.$
As the quadratic term is always positive in the $\delta$-PT2 approximation the ensemble 
correlation energy becomes positive.
In particular if the density $n = 1.5$, the following expression is obtained:
\begin{eqnarray}
e_{c}^{w=0.5,\delta-PT2}(u) = \frac{1}{4} - \frac{3}{8u} + \mathcal{O}\left(\frac{1}{u^2}\right).
\end{eqnarray}
Consequently as observed with the Fig. (\ref{fig:positivecorrelationn15}) an asymptote is reached 
for $e_{c}^{w,\delta-PT2}(n) = 0.25.$
However if the density is very close to $1.5$ such that $n = 1.5-\eta$ with $\eta > 0$
a negative slope is given in the expansion and the correlation energy can become negative
for very large value for $U.$
\fi%%%%%%

%%%%%% Fig. %%%%%%%%%%%%%%%%%%%

%%%%%%%%%%%%%%%%%%%%%%%%%%%%

\begin{figure*}
\resizebox{1\textwidth}{!}{
\includegraphics[scale=1]{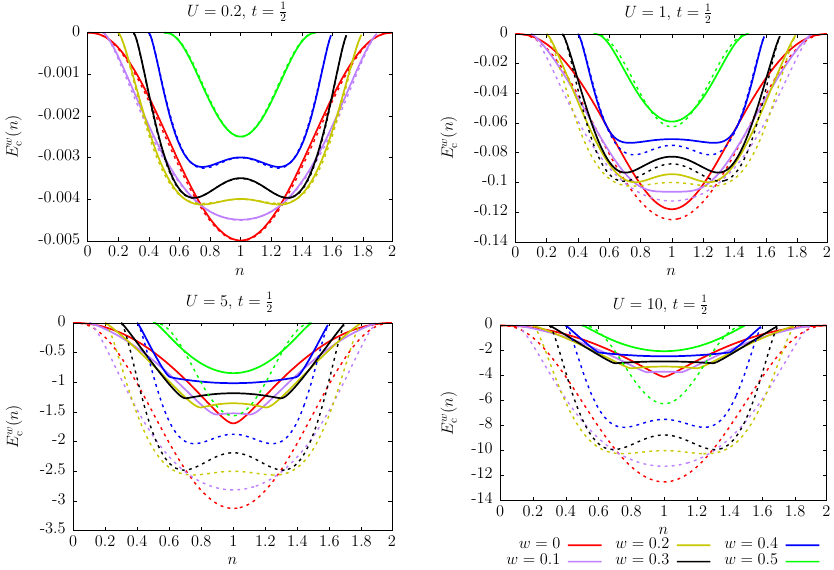}
}
\caption{$u$-PT2 correlation energy (dashed lines) plotted as a function of the density
for various correlation regimes and ensemble weights. Comparison is made with the exact
results (solid lines) of Ref.~\cite{deur2017exact}.}
\label{fig:EcuPT2}
\end{figure*}

Turning to the $u$-PT2 approximation (see Eq.~(\ref{eq:Taylor_u_second_order}) and Fig.~\ref{fig:EcuPT2}),
accurate correlation energies are obtained in the weakly correlated
regime for {\it all} densities, as expected. Errors become large,
especially around the symmetric $n$=1 ensemble density, as $U$
increases. Interestingly, the equi-ensemble seems to
be less affected by the overestimation of the correlation energy than
ensembles where the ground state dominates (i.e. $w \ll 1/2$). 
Finally, unlike $\delta$-PT2, $u$-PT2 gives by
construction (see Eq.~(\ref{eq:Taylor_u_second_order})) the correct
correlation energy [which is equal to zero as shown in Appendix~\ref{appendix:Ec_border}] at the border of the
representability domain. 
%In the rest of the paper, approximate calculations will be performed for equi-ensembles only.

\subsection{
Density-functional total energy profile and minimizing 
densities for equi-ensembles}\label{subsec:energy_profile}

\begin{figure*}
\resizebox{1\textwidth}{!}{
\includegraphics[scale=1]{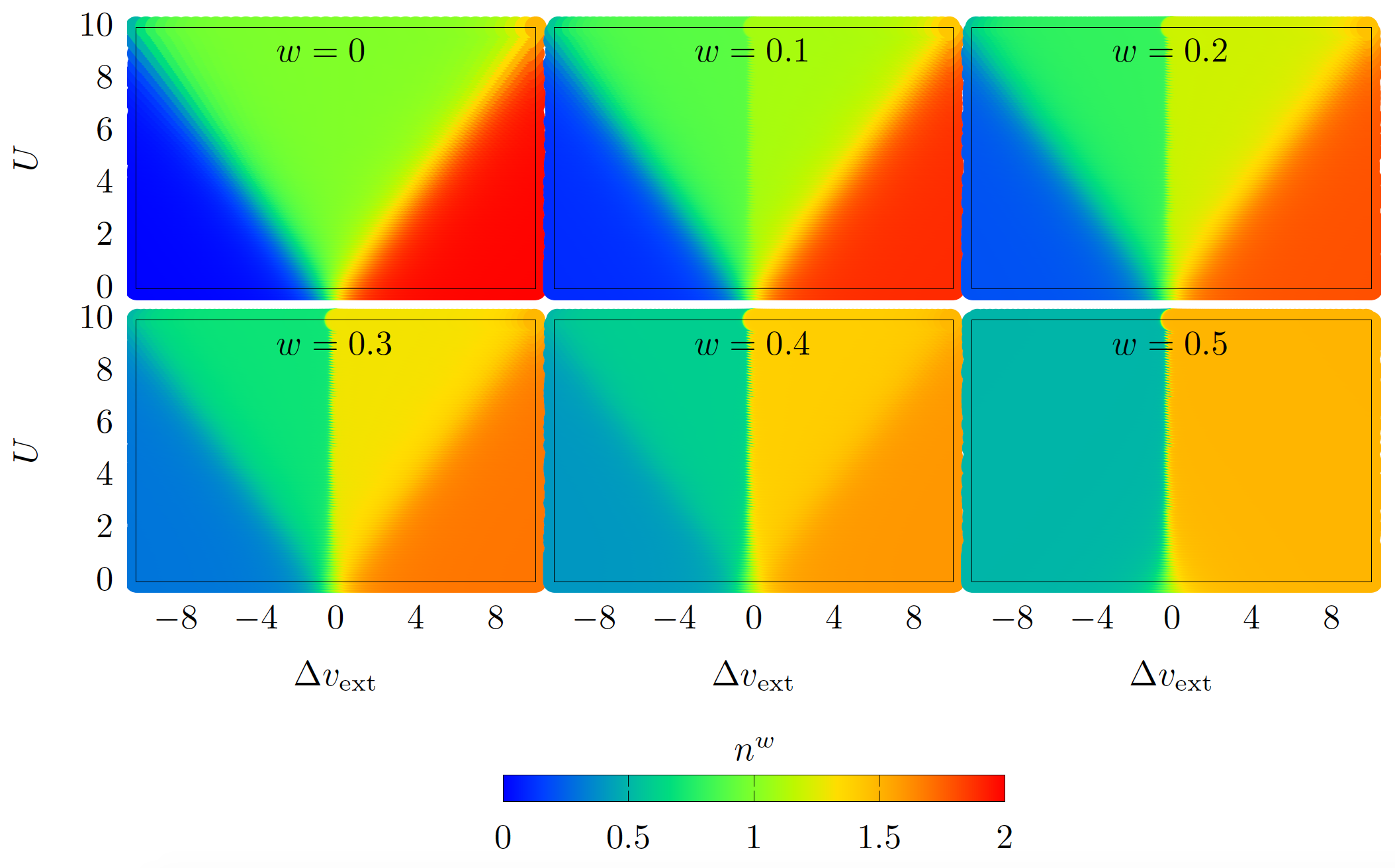}
}
\caption{Map of the exact ensemble density plotted as a function of $U$
and $\Delta v_{\rm ext}$ for various ensemble weights.}
\label{fig:ensemble_dens}
\end{figure*}

This section deals with the optimization of the ensemble density through
minimization of the total ensemble density-functional energy in
Eq.~(\ref{eq:densfun_Ew_GOKDFT}). For analysis purposes, exact ensemble
densities have been plotted in Fig.~\ref{fig:ensemble_dens} with respect to $U$ and
$\Delta v_{\rm ext}$ for various values of the ensemble weight $w$. As
expected from Ref.~\cite{deur2017exact}, the interacting density profile ($U > 0$) 
satisfies the non-interacting $v$-representability condition in
Eq.~(\ref{eq:non_int_vrep_cond}). Density domains can clearly be
distinguished and, in particular, it appears that the ensemble density undertakes critical changes around $U / \Delta v_{\rm ext} \approx \pm 1$ and $\Delta v_{\rm ext} 
\approx 0$, which can be summarized as follows when $U/(2t)$ is
sufficiently large,
\begin{eqnarray}\label{eq:ens_dens_discontinuities}
n^w \approx \left\lbrace \begin{array}{lcr}
w && -1 < \dfrac{U}{\Delta v_{\rm ext}} < 0 \\ \\
1-w & & -\infty < \dfrac{U}{\Delta v_{\rm ext}} < -1 \\ \\
1& \text{~~~for~~~} & \Delta v_{\rm ext} \approx 0\\ \\
1+w && +1 < \dfrac{U}{\Delta v_{\rm ext}} < +\infty\\ \\
2-w &  & 0 < \dfrac{U}{\Delta v_{\rm ext}} < +1.\\
\end{array}
\right.
\end{eqnarray}
Note that, in the particular case of the equi-ensemble ($w=1/2$), ensemble
densities will essentially be equal to 1/2, 1 (in the vicinity of the
symmetric case) or 3/2.\\
%\subsection{convergence issues}\label{subsec:conv}

Let us now focus on the approximate calculation of ensemble
densities. Calculating the ensemble energy profiles for the set of
non-interacting $v$-representable ensemble densities within all
aforementioned approximations will allow us to detect possible local
minima that can lead to wrong minimizing ensemble densities and
convergence issues. In the exact theory, both non-interacting kinetic and xc
functionals are weight-dependent so that the total density-functional
energy is strictly convex. In practical calculations, however, there is
no straightforward way to develop weight-dependent functionals and one
has to recur to approximations such as neglecting the weight dependence,
like in GSxc (see Sec.~\ref{sec:comput_details}). In the following, we discuss what effect the neglect or
the (partial) introduction of weight dependence in the xc functional has
on the profile of the total ensemble energy.\\

\begin{figure}
\resizebox{0.49\textwidth}{!}{
\includegraphics[scale=1]{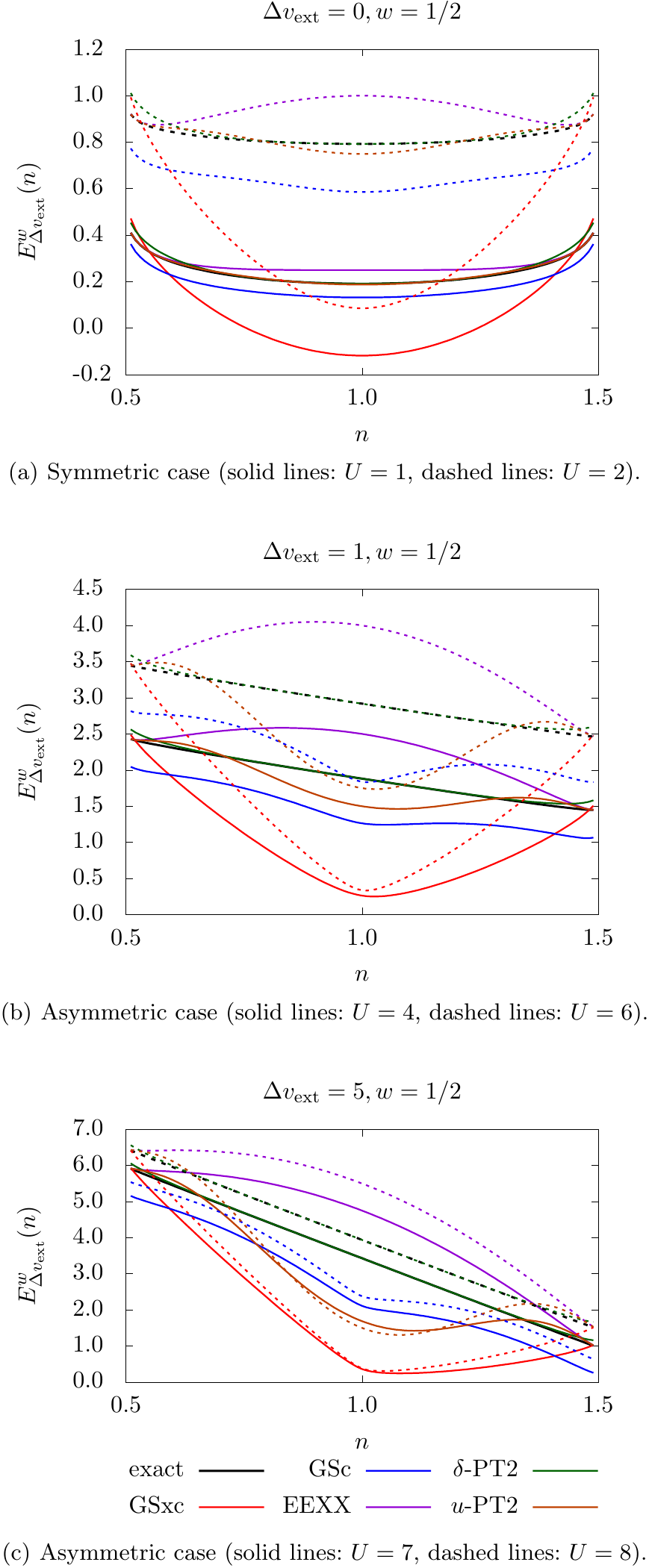}
}
\caption{Exact and approximate total equi-ensemble density-functional
energies plotted 
for various interaction strengths and external
potentials. See text for further details.}
\label{fig:ensemble_energy_profile}
\end{figure}

\begin{figure}
\resizebox{0.49\textwidth}{!}{
\includegraphics[scale=1]{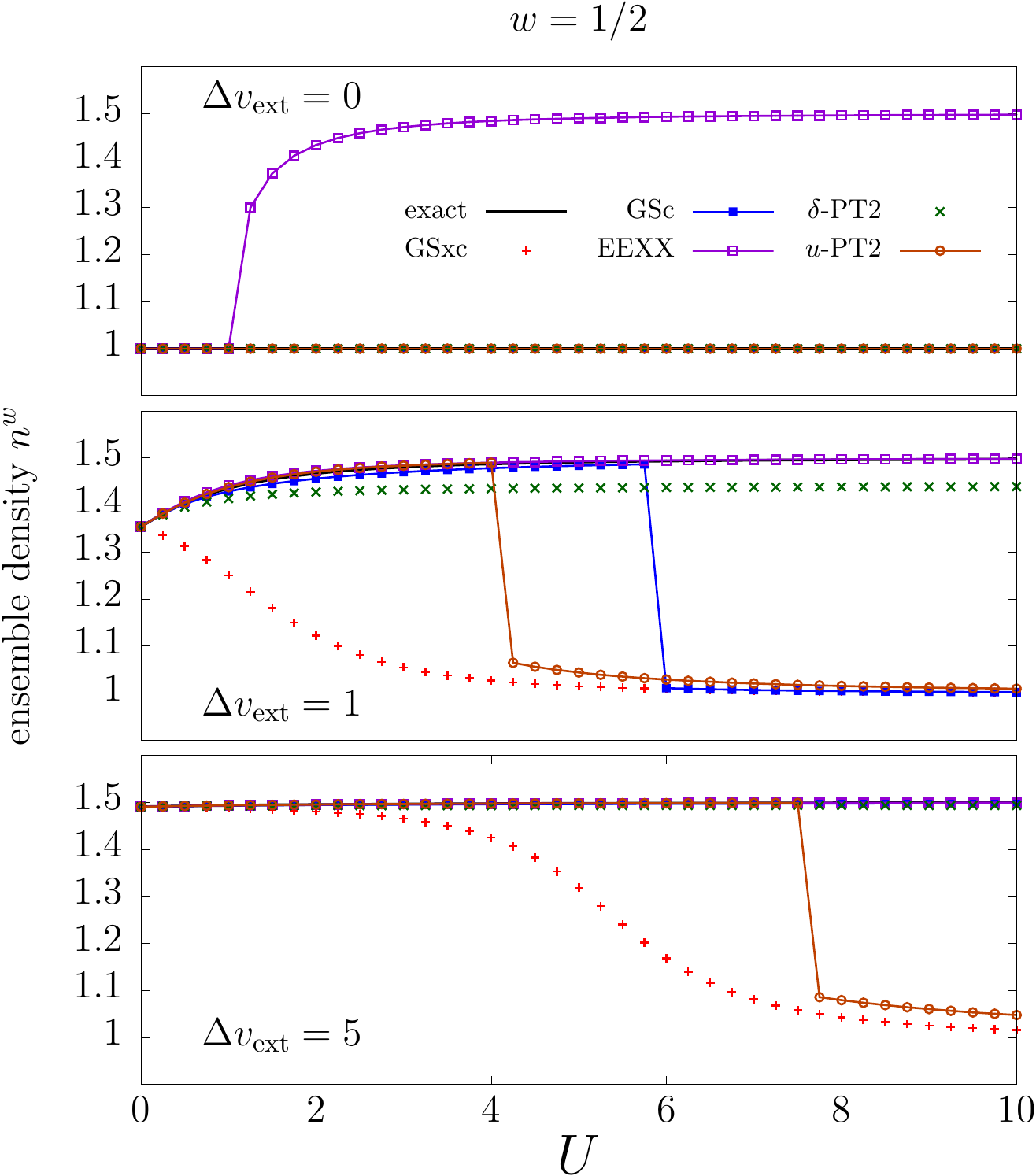}
}
\caption{Exact and approximate equi-ensemble minimizing densities plotted as a 
function of $U$ for various external potentials. See text for further
details. 
}
\label{fig:nmin_deltav_w05}
\end{figure}

A selection of peculiar and
problematic cases are plotted in Fig. \ref{fig:ensemble_energy_profile}.
The minimizing ensemble densities are obtained by global brute-force
minimization and are plotted as a function of $U$ for the symmetric and
asymmetric cases in Fig. \ref{fig:nmin_deltav_w05}.
The GSxc approximation has no convexity issue as both the kinetic and
the exact ground-state functionals are strictly convex.
%Convergence
%issues may arise in the strongly correlated regime because of the
%discontinuity in the ground-state xc potential at
%$n=1$~\cite{carrascal2015hubbard,deur2017exact}.
Nevertheless, it gives quite poor equi--ensemble energies (see Fig.
\ref{fig:ensemble_energy_profile}), which is due to the fact that, for
$w=1/2$, the excited state contributes to half of the ensemble energy
and thus the weight dependence cannot be completely neglected. The
minimizing ensemble densities are correct in the symmetric case (top
panel of Fig. \ref{fig:ensemble_energy_profile}) but as soon as $\Delta
v_{\rm ext}$ increases they are too far off from the exact ones (see
Fig. \ref{fig:nmin_deltav_w05}). Turning to the GSc approximation (see Sec.~\ref{sec:comput_details}), the equi--ensemble energy profile is
not strictly convex for all values of $U$ (see Fig.
\ref{fig:ensemble_energy_profile}). On the one hand, adding the EEXX
to the ground-state correlation functional yields better equi-ensemble
energies than GSxc but they are still too poorly described. The
minimizing ensemble densities, on the other hand, are exact in the
symmetric case and in asymmetric cases where  $\Delta v_{\rm ext}$ is
sufficiently large compared to $U$ (see Fig. \ref{fig:nmin_deltav_w05}).
However, in the intermediate case, i.e. when $U\gg \Delta v_{\rm ext}$
(see the middle panel of Fig. \ref{fig:ensemble_energy_profile}), the
global minimum abruptly switches place with another minimum located at
$n=1$ and causes the discontinuity at $U=6$ in the plot of the minimizing ensemble
density as a function of $U$ (see the middle panel of Fig. \ref{fig:nmin_deltav_w05}). 
Let us stress that, even when GSc gives the right density by
global minimization, the existence
of local minima and maxima in the strongly correlated regime will lead to serious convergence issues when
searching for stationary densities, which would be equivalent to solving
the ensemble KS equations self-consistently. This is due not only to the    
discontinuity in the ground-state xc potential at
$n=1$~\cite{deur2017exact,carrascal2015hubbard} but also to the
non-convexity of the equi-ensemble energy profile induced by the
complete neglect of weight dependence in the correlation energy
contribution.\\ 

Neglecting the correlation energy in the GSc scheme leads to the EEXX
approximation. In the latter case, the minimization can be carried out
analytically for the symmetric dimer (see Appendix
\ref{appendix:eexx_min}). When $w\leq 1/3$, the ensemble energy has a
unique minimizing ensemble density, whereas for $w>1/3$ there is a
critical value of $U$ beyond which the strict convexity is suppressed
and two degenerate minima appear on the ensemble energy profile (see the top
panel of Fig. \ref{fig:ensemble_energy_profile}). In the specific case
where $w=1/2$, this value is $U=1$. This abrupt change explains why the
EEXX minimizing ensemble density exhibits a 
discontinuity in the top panel of Fig. \ref{fig:nmin_deltav_w05}. 
%The same problem is inherited by the GSc approximation (middle panel of Fig. \ref{fig:nmin_deltav_w05}). 
Away from the symmetric case, the EEXX equi-ensemble energy has the 
correct global minimum even though it exhibits non-convexity. 
Note that, as shown in Appendix~\ref{appendix:Ec_border}, the equi-ensemble EEXX energy is exact at the border of the
non-interacting $v$-representability domain, i.e. when $n=1/2$ or
$n=3/2$.\\  

Turning to the weight-dependent $\delta$-PT2 correlation DFA (see
Eq.~(\ref{eq:Taylor_exp_2nd_order_delta})), the
equi-ensemble energy exhibits convexity in both weakly and strongly
correlated regimes (see Fig. \ref{fig:ensemble_energy_profile}). Thus,
unlike GSc and EEXX, the minimization scheme is robust and does not lead
to discontinuities in the minimizing ensemble densities. $\delta$-PT2 is
essentially exact around the symmetric case, by construction. 
Errors appear in the minimizing density when $U\gg \Delta v_{\rm ext}>0$
(see the middle panel of Fig. \ref{fig:nmin_deltav_w05}). As in GSc, as
soon as $\Delta v_{\rm ext}$ is sufficiently large, the exact minimizing
ensemble density is almost recovered. Moreover, thanks to the absence of
density
derivative discontinuities in the $\delta$-PT2 correlation functional
(see Fig. \ref{fig:EcdeltaPT2}), self-consistent calculations of
(stationary) ensemble densities in
the strongly correlated regime are
expected to converge smoothly, which is clearly an advantage from a
practical point of view.\\ 

Let us finally discuss the performance of the $u$-PT2 approximation
which uses a weight dependent density-functional correlation energy based on a
perturbative expansion of the exact correlation energy around $U=0$ (see
Eq.~(\ref{eq:Taylor_u_second_order})). As
expected, $u$-PT2 performs well in the weakly correlated regime. As soon
as $U$ increases, it faces the same problem as Gsc and EEXX (see middle
and bottom panels of Fig. \ref{fig:ensemble_energy_profile}). Indeed,
the ensemble energy loses convexity (local minima and maxima appear), which leads to discontinuities in the minimizing ensemble densities (see
the middle and bottom panels of Fig. \ref{fig:nmin_deltav_w05}).\\

In summary, including weight dependence into the ensemble correlation energy
is crucial in order to obtain quantitatively good results
(densities and energies) and avoid potential convergence issues when
searching for stationary densities of the total energy or, equivalently,
when solving the self-consistent ensemble KS equations. Furthermore, keeping only the weight dependence in the exchange part has proven to be insufficient.
The best reproduction of the exact equi-ensemble energy profiles and minimizing ensemble densities is by far obtained by the $\delta$-PT2 approximation. It is valid for both the weakly and strongly correlated regime and, despite being based on an expansion around $n=1$, it also yields decent results in the asymmetric case.

\iffalse%%%%%%
The GSxc approximation leads to strictly convex equi-ensemble energy profiles, but is not suited to describe the excited state. The GSc,  EEXX and $u$-PT2 approximations only reproduce well the minimizing ensemble densities for certain cases and the equi--ensemble energy profiles exhibit bumps and local minima which makes minimization and SCF calculations unstable. Nevertheless, these partially weight-dependent approximations may yield excellent results through error cancellation between the kinetic and Hartre-excange part when calculating excitation energies via LIM. 
\fi%%%%%%%%%%%%

\iffalse%%%%%%%%%%% read and commented by Manu %%%%%%%%%%%
\textcolor{red}{Bruno : should be said in Laurent's part somehow:
First of all, note that a minimizing density is always obtained by 
construction, in contrast to the usual self-consistent procedure
where discontinuity(ies) in the potential sometimes prevent from 
reaching
a stationary point.
The minimization, by avoiding the implementation of the potential,
is therefore more robust than the self-consistent procedure,
but one has to compute the ensemble energy for every density in
the ensemble $v$-representable domain $w < n < 2-w$.
If it happens that two minima are degenerated, there is always one 
in between $w$ and $1$ and one in 
between $1$ and $2-w$. The correct 
minima is selected depending on the sign of $\Delta v_{\rm ext}$.
If $\Delta v_{\rm ext} < 0$, then $n^w < 1$ and vice-versa.}
\fi%%%%%%%%%%%%%%%

\iffalse
\subsection{ensemble energies}

The ensemble energy is plotted in Figs.~(\ref{fig:Ew_deltav1}) and (\ref{fig:Ew_deltav5}).

\begin{figure}
\resizebox{0.49\textwidth}{!}{
\includegraphics[scale=1]{../figures/Ew_deltav1_fctU_fctw.pdf}
}
\caption{Ensemble energy for $U=1$ (top panel) $U=5$ (middle panel) and $U=10$ (bottom panel) at iteration 0 (full lines) and after convergence (dashed lines) for $\Delta v = 1$. If dashed lines are missing, it means that the convergence was not reached.}
\label{fig:Ew_deltav1}
\end{figure}

\begin{figure}
\resizebox{0.49\textwidth}{!}{
\includegraphics[scale=1]{../figures/Ew_deltav5_fctU_fctw.pdf}
}
\caption{Ensemble energy for $U=1$ (top panel) $U=5$ (middle panel) and $U=10$ (bottom panel) at iteration 0 (full lines) and after convergence (dashed lines) for $\Delta v = 5$. If dashed lines are missing, it means that the convergence was not reached.}
\label{fig:Ew_deltav5}
\end{figure}
\fi%%%%%%%%%%%%%%%%%%%%%

% Bruno : I need this figure, please don't remove it.
%%%%%%%%%%%%%%%%%%%%%%%%%%%%%%%

\subsection{Ensemble energy derivatives}

\begin{figure}
\resizebox{0.49\textwidth}{!}{
\includegraphics[scale=1]{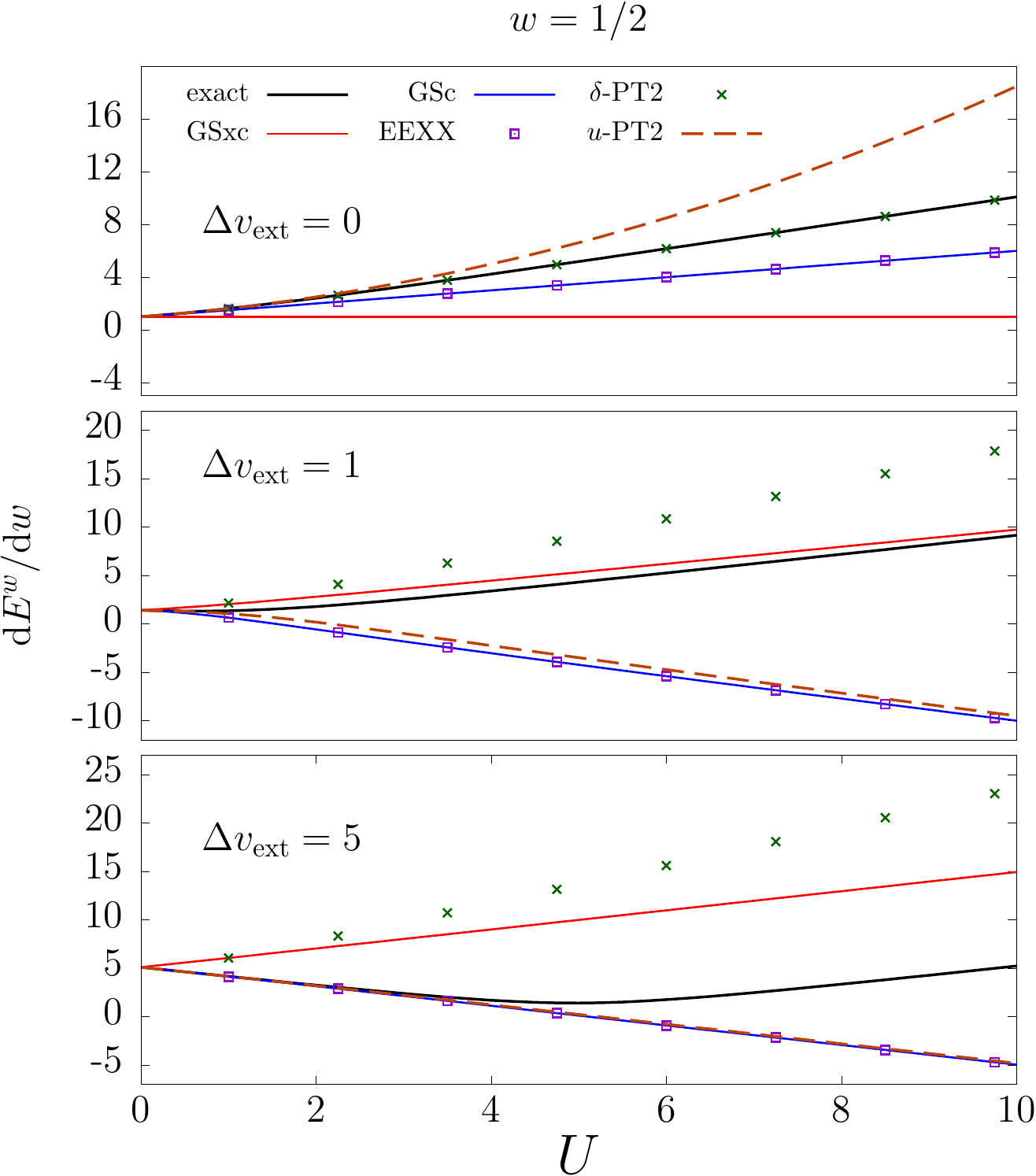}
}
\caption{
Exact and approximate ensemble energy derivatives calculated
with 
the {\it exact} ensemble density and plotted as a
function of $U$ in the equi-ensemble case and for various external
potentials. 
}
\label{fig:dEw_dw_nexact}
\end{figure}

\begin{figure}
\resizebox{0.49\textwidth}{!}{
\includegraphics[scale=1]{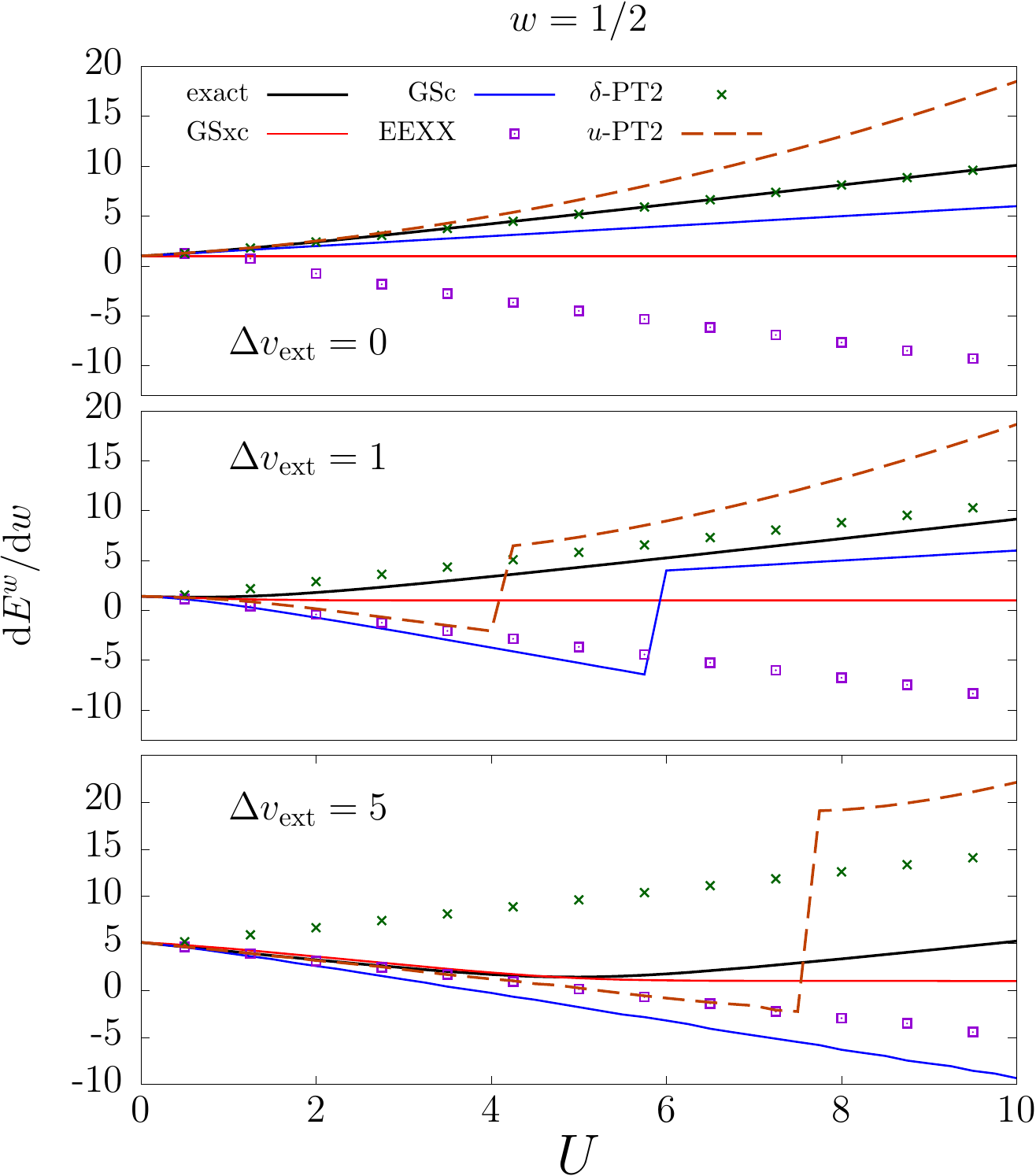}
}
\caption{
Exact and approximate ensemble energy derivatives calculated
with 
the {\it minimizing} ensemble densities and plotted as a
function of $U$ in the equi-ensemble case and for various external
potentials. 
}
\label{fig:dEw_dw_nmin}
\end{figure}

\begin{figure}
\resizebox{0.49\textwidth}{!}{
\includegraphics[scale=1]{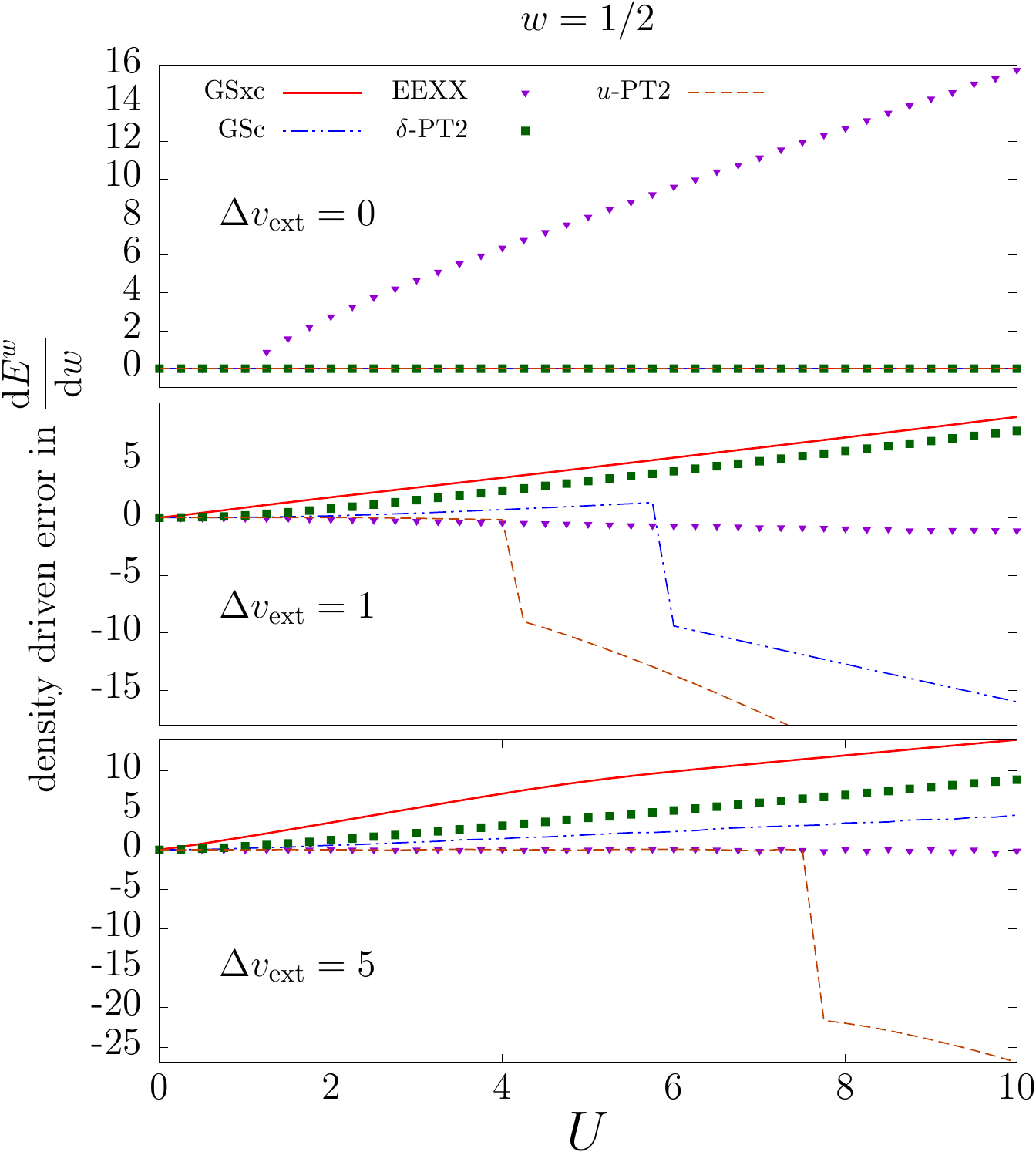}
}
\caption{density driven error in the ensemble energy derivative
evaluated for each DFA as the difference between 
$
\left.
%\dfrac
{\ddroit E_{\rm DFA}^w(n)}/{\ddroit w}\right|_{n=n^w_{\rm exact}}
$
and 
$\left.
%\dfrac
{\ddroit E_{\rm DFA}^w(n)}/{\ddroit w}\right|_{n=n^w_{\rm DFA}}
$, where $n^w_{\rm DFA}$ denotes the minimizing ensemble density. The
result is plotted as a function of $U$ in the 
equi-ensemble case and for various external
potentials. 
}
\label{fig:density_driven_error_dEwdw}
\end{figure}

In practice, any weight in the range $0 \leq w \leq 1/2$ can in principle be used
for computing the excitation energy. As argued in
Sec.~\ref{subsec:testing_DFA}, we expect the equi-ensemble case ($w =
1/2$) to be the most favorable one for the DFAs discussed previously,
especially $\delta$-PT2 (see Eq.~(\ref{eq:Taylor_exp_2nd_order_delta})). We focus in this section on the calculation of
approximate excitation energies by differentiation (see Eq.~(\ref{eq:dEw_dw_HD})). 
In order to evaluate both functional driven and total errors, results
obtained with the exact and the minimizing ensemble densities are shown   
in Figs.~\ref{fig:dEw_dw_nexact} and~\ref{fig:dEw_dw_nmin},
respectively. In addition, the difference between the two excitation
energies is plotted in Fig.~\ref{fig:density_driven_error_dEwdw}, in
order to visualize the impact of density driven errors. 

%%%%%%%%%%%%%%%%%%%%%%%%%%%%%%%%%%%%%%%%%%%%%
%%%%%%%%%%%%%%%%%%%%%%%%%%%%%%%%%%%%%%%%%%%%%
%%%%%%%%%%%%%%%%%%%%%%%%%%%%%%%%%%%%%%%%%%%%%

As shown in Fig.~\ref{fig:dEw_dw_nmin}, all the DFAs using 
a weight independent ensemble correlation energy (namely GSxc, GSc, and EEXX) underestimate the
excitation energy. Unphysical negative excitation energies are even
obtained with GSc and EEXX in the strongly correlated regime, as
expected~\cite{deur2017exact}. In the latter regime, EEXX exhibits large density driven
errors in the symmetric case only, in agreement with Sec.~\ref{subsec:energy_profile}. 
In the asymmetric case, the error is
purely functional driven. The opposite is observed for GSxc. Note that,
at the GSc level of approximation and for $\Delta v_{\rm ext}=1$ (see
the middle panel of Fig.~\ref{fig:dEw_dw_nmin}), the excitation energy
exhibits a discontinuity around $U=6$, as expected from
Sec.~\ref{subsec:energy_profile}. Interestingly, even though GSc gives a
completely wrong ensemble density in this regime of correlation, the
accumulation of functional and density driven errors (see the middle
panels of Figs.~\ref{fig:dEw_dw_nexact} and \ref{fig:density_driven_error_dEwdw}) leads to 
relatively good excitation energies.\\

Turning to weight-dependent correlation DFAs,  
$u$-PT2 (see Eq.~(\ref{eq:Taylor_u_second_order})) performs well only for relatively small $U$ values, as expected.
The discontinuities observed for large $U$ values in asymmetric cases
are induced by sudden changes in the minimizing ensemble density as $U$
increases (see Sec.~\ref{subsec:energy_profile} for further details).  
Unlike GSc, $u$-PT2 does not benefit from error cancellations in the strongly
correlated regime. In the asymmetric case, the excitation energies are
indeed significantly overestimated (see the middle and bottom panels of
Fig.~\ref{fig:dEw_dw_nmin}). Note that taking into account functional
driven errors only would lead to negative excitation energies in this
case (see the middle and bottom panels of
Fig.~\ref{fig:dEw_dw_nexact}). Thanks to (too) large additional density driven errors
(see the middle and bottom panels in Fig.~\ref{fig:density_driven_error_dEwdw}),
positive excitation energies are finally obtained.\\  

Let us now focus on the $\delta$-PT2 approximation (see
Eq.~(\ref{eq:Taylor_exp_2nd_order_delta})). It is, by
construction, exact for the symmetric dimer. In the asymmetric $\Delta v_{\rm ext} = 5$ case,
however, $\delta$-PT2 overestimates the excitation energy significantly
as $U$ increases. This was actually expected from
Eqs.~(\ref{eq:DD_0.5_minus_u_infty}) and~(\ref{eq:DD_0.5_plus_u_infty})
since the ensemble density is, in this case, close to 3/2 (see the
bottom panel of Fig.~\ref{fig:nmin_deltav_w05}). Interestingly, the
density driven error is substantial in this case (see the bottom panel of
Fig.~\ref{fig:density_driven_error_dEwdw}), which is quite surprising as
minimizing and exact densities are very similar. As readily seen from
Eq.~(\ref{eq:dEw_dw_HD}), the non-interacting kinetic energy
contribution to the ensemble energy derivative has a singularity at
$n=3/2$, thus making the excitation energy highly sensitive 
to changes in the density. Note finally that, even though the
$\delta$-PT2 excitation energy is too high in this regime of density and
correlation, the density driven error
removes a significant part of the functional driven one. 
\\

\subsection{Linear interpolation method}

\begin{figure}
\resizebox{0.49\textwidth}{!}{
\includegraphics[scale=1]{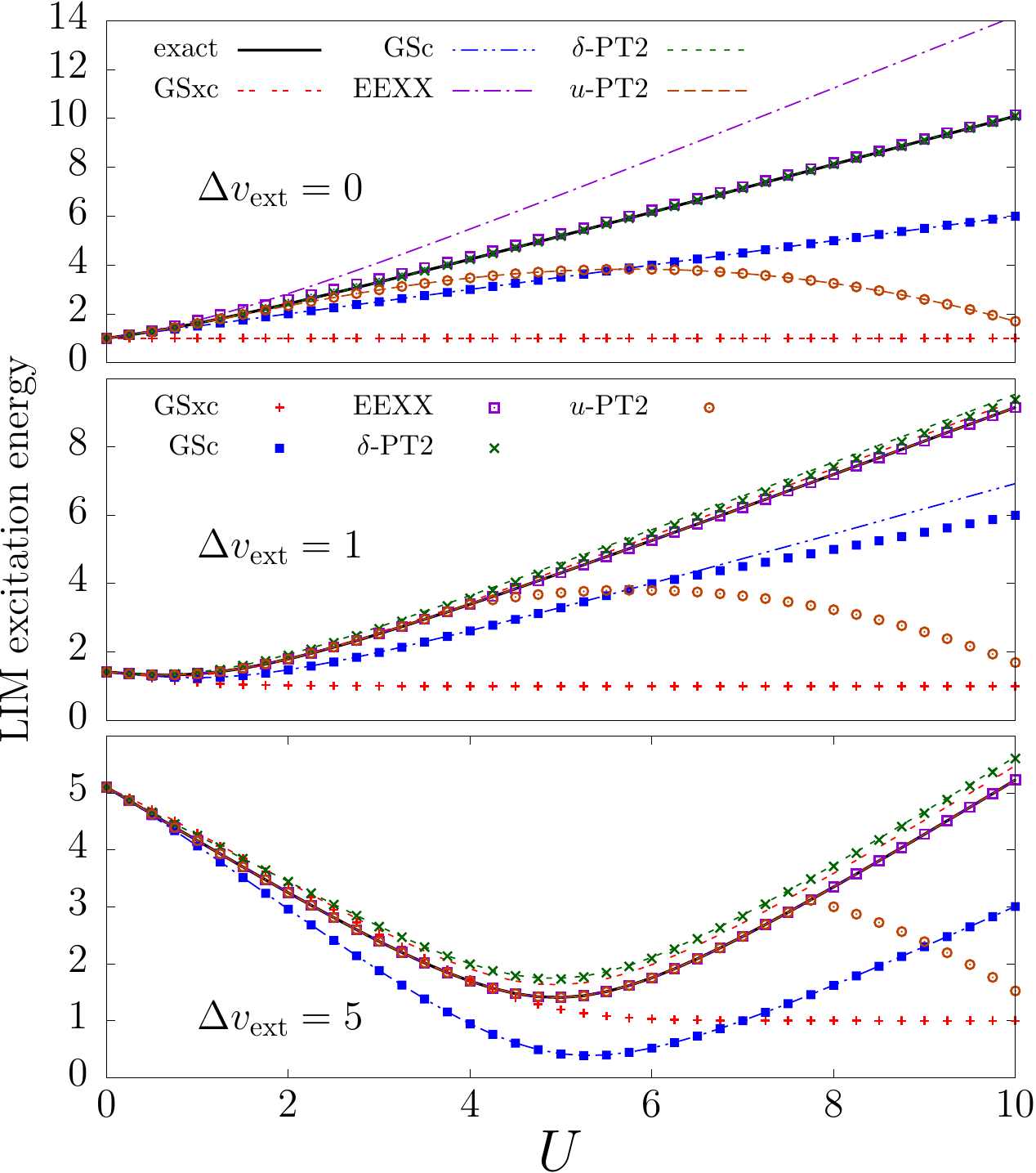}
}
\caption{Excitation energies calculated with respect to $U$ by linear interpolation for various DFAs and
external potentials. Results obtained with the {\it exact} equi-ensemble
density (dashed lines) are compared with those obtained with the
minimizing densities (shown with points). Colors are used for
distinguishing the DFAs. 
}
\label{fig:LIM}
\end{figure}

\begin{figure}
\resizebox{0.49\textwidth}{!}{
\includegraphics[scale=1]{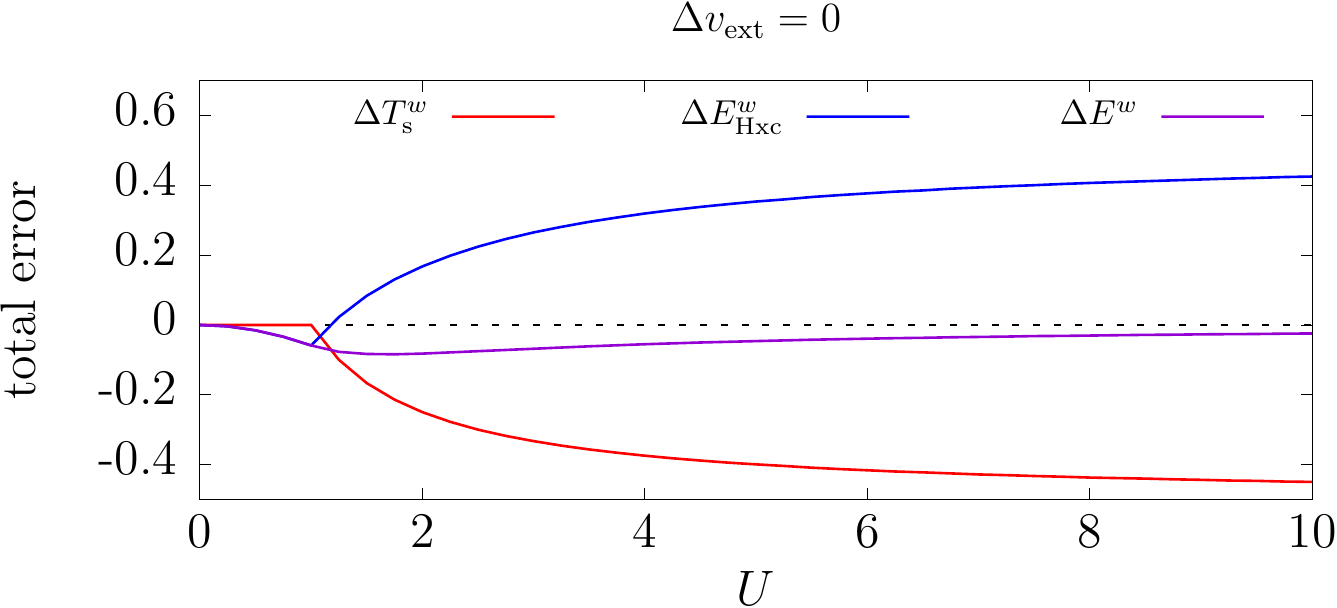}
}
\caption{Total error $\Delta E^w = \Delta T_{\rm s}^w + \Delta E_{\rm Hxc}^w$ in the equi-ensemble ($w=1/2$) energy plotted as a
function of $U$ for the EEXX approximation in the symmetric dimer.
Non-interacting kinetic energy 
$\Delta T_{\rm s}^w = T_{\rm s}^w(n^w_{\rm exact}) - 
T_{\rm s}^w(n^w_{\rm EEXX})$ 
and Hxc energy
$\Delta E_{\rm Hxc}^
{w}
 = E
^
{w,{\rm exact}}
_{\rm Hxc}
(n^w_{\rm exact}) - 
E_{\rm Hx}
^
{w,{\rm exact}}
(n^w_{\rm EEXX})$ contributions are shown in order to
highlight error cancellations. 
}
\label{fig:EEXX_comp_error}
\end{figure}

The linear interpolation method (LIM)~\cite{senjean2015linear} is an alternative to the differentiation of the ensemble energy
for the extraction of 
excitation energies. As readily seen from Eq.~(\ref{eq:LIM_exact}), the
latter are calculated within LIM from 
both
ground-state and equi-ensemble energies. Since we use the accurate
parameterization of Carrascal {\it et al.}~\cite{carrascal2015hubbard,carrascal2016corrigendum}
for the ground-state correlation functional, errors in our LIM
excitation energies will exclusively originate from the ensemble xc DFA
that is used.  
Results obtained with the exact and minimizing ensemble densities are shown in Fig.~\ref{fig:LIM}.
In the symmetric case (top panel), $\delta$-PT2 (see Eq.~(\ref{eq:Taylor_exp_2nd_order_delta})) is exact in all
correlation regimes while $u$-PT2 (see Eq.~(\ref{eq:Taylor_u_second_order})) performs well only for relatively
small $U$ values, as expected. The lack of weight dependence in GSxc and
GSc leads to an underestimation of the excitation energy. 
EEXX performs surprizingly well in this case, even though it exhibits large
functional driven {\it and} density driven errors. As shown in
Fig.~\ref{fig:EEXX_comp_error}, these errors cancel each other as $U$
increases. Note that the interaction derivative discontinuities around $U = 1$ in the non-interacting and Hxc
ensemble energies originate from the sudden change in the minimizing
ensemble density discussed previously (see the top panel of Fig.~ \ref{fig:nmin_deltav_w05}).
\\

\begin{figure}
\resizebox{0.49\textwidth}{!}{
\includegraphics[scale=1]{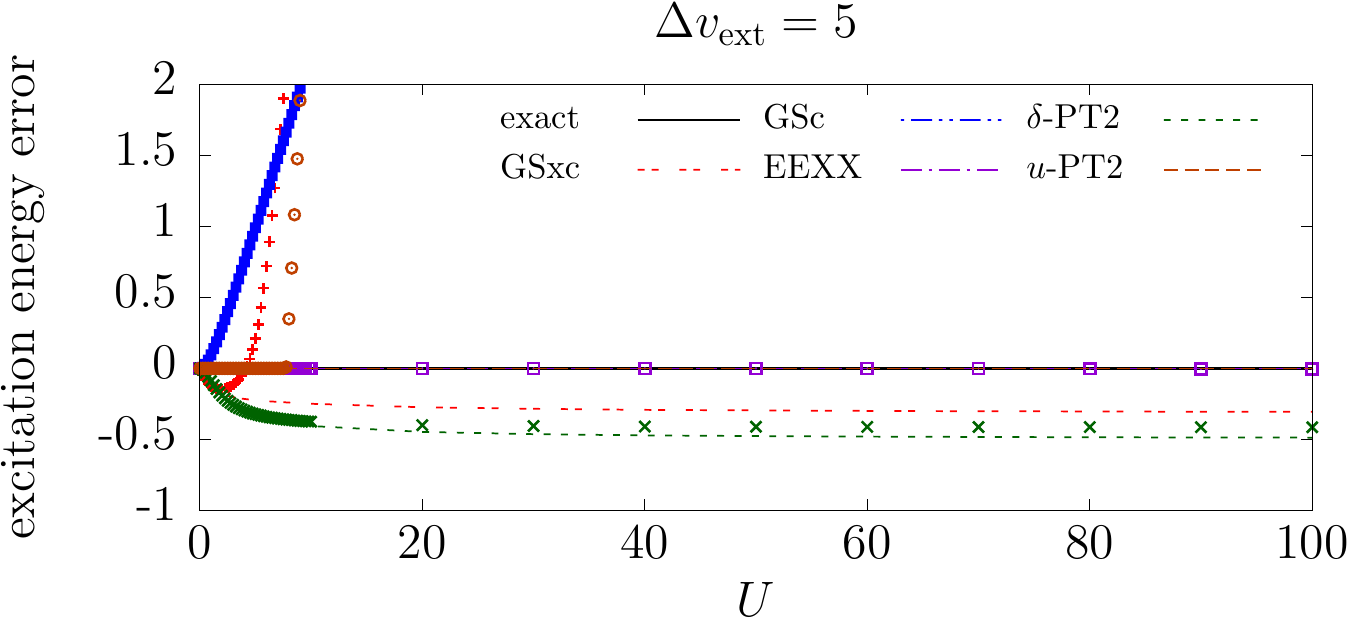}
}
\caption{Errors in the LIM excitation energy plotted as a function of $U$
for the various DFAs and  
$\Delta v_{\rm ext} = 5$. The total error [$\Omega^{{\rm exact}}-\Omega^{\rm
LIM}(n^{w=1/2}_{\rm DFA})$] is shown with points. Dashed lines are used for
plotting the functional driven error contribution [$\Omega^{{\rm exact}}-\Omega^{\rm
LIM}(n^{w=1/2}_{\rm exact})$].
Colors are used for
distinguishing the DFAs. See text for further details.
}
\label{fig:LIM_U100}
\end{figure}

Turning to asymmetric cases (see the middle and bottom panels of 
Fig.~\ref{fig:LIM}), EEXX and exact excitation energy curves are
essentially on top of each other. This is simply due to the fact that, as $U$ increases, the ensemble
density becomes very close to 3/2 so that the
equi-ensemble correlation energy vanishes (see Appendix~\ref{appendix:Ec_border}). Note that, at the EEXX level
of approximation, both functional and density driven errors are equal to zero in this case.  
Interestingly, inserting the exact equi-ensemble density into the GSxc
functional gives 
relatively good results in the strongly correlated regime, as
expected~\cite{deur2017exact}. However, as shown in
Fig.~\ref{fig:LIM_U100}, large density driven errors lead to a
significant underestimation of the excitation energy in this regime.   
Note finally that, as expected from Ref.~\cite{deur2017exact}, GSc 
systematically underestimates the excitation energy.
Regarding the weight dependent correlation DFAs, $u$-PT2 performs
as well as EEXX when the exact ensemble density (which is close to 3/2)
is used, as expected from Eq.~(\ref{eq:Taylor_u_second_order}). Unlike
EEXX, $u$-PT2 suffers from significant density driven errors (see
Fig.~\ref{fig:LIM_U100}) when $U$ is sufficiently large, thus leading to
a deterioration of the excitation energy. On the other hand,
$\delta$-PT2 gives relatively accurate results in the same regime of
density and correlation. As shown in Fig.~\ref{fig:LIM_U100} and expected from Eqs.~(\ref{eq:LIM_exact}),
(\ref{eq:deltaPT2_0.5_minus_eta_u_infty}) and
(\ref{eq:exp_u_infty_exact_delta_0.5}), a residual error (equal
to -0.5 per unit of $2t$ when exact densities are used) is obtained as $U$
increases, which is due to an unphysical positive correlation energy
contribution. Note that the latter error is essentially functional
driven.

\section{Conclusions}\label{sec:Conclusion}

Ensemble DFT for excited states has been applied to the two-electron Hubbard
dimer. An ensemble consisting of the first two singlet states has been
considered. While Ref.~\cite{deur2017exact}  was focusing on the exact calculation of
(weight-dependent) correlation energies, the design of
analytical density-functional approximations (DFAs) as well as their performance
in practical (self-consistent) calculations has been investigated in
this work. Simple DFAs based on the expansion of the exact ensemble
Hohenberg--Kohn functional in various density and correlation regimes
have been considered. Note that the use of Legendre--Fenchel transforms (rather
than the usual Levy--Lieb constrained-search formalism which would
involve two many-body wavefunctions) is extremely convenient since a single variable, namely the local
potential difference $\Delta v$, is needed (see
Eq.~(\ref{universalfunctionalHD})).\\

As clearly shown in this
simple but nontrivial model, the equi-ensemble case is the simplest one
to model, as long as the ensemble density remains away from the border
of the non-interacting $v$-representability domain.
If so, the most reliable DFA,
referred to as $\delta$-PT2, uses an expansion through second order in
the density deviation $\delta=n-1$ from the symmetric case (see
Eq.~(\ref{eq:Taylor_exp_2nd_order_delta})). It performs
relatively well in both weakly and strongly correlated regimes and is,
by construction, exact for the symmetric dimer.\\

Excitation energies have
been extracted in two ways. The differentiation with respect to the ensemble
weight (see Eq.~(\ref{eq:XE_deriv_ab_init})) does not always give satisfactory results, especially in the
asymmetric strongly correlated regime. A proper description of the
weight-dependent correlation density derivative discontinuities in the strongly
correlated regime would be needed.
Those actually appear at the border of the density domain of
applicability of 
$\delta$-PT2. The second approach (see Eq.~(\ref{eq:LIM_exact})), namely the linear
interpolation method (LIM), is much more reliable especially since it avoids
the difficult task of modelling the xc derivative discontinuity. Despite a spurious
positive correlation energy contribution which appears in the asymmetric case as the on-site repulsion
increases, combining $\delta$-PT2 with LIM gives relatively accurate results.                 
Note that this spurious contribution originates from the fact that the equi-ensemble
density is outside the domain of applicability of $\delta$-PT2. Better
DFAs would be obtained by interpolating the expansions obtained in the
various regimes. This is left for future work.\\

Let us finally stress that the combination of   
the ensemble exact exchange-only energy (EEXX) [see Eq.~(\ref{eq:EEXX_fun_HD})] with LIM yields remarkably accurate
excitation energies, even in the strongly correlated regime. In the
symmetric case, both functional and density driven errors are large  
but they actually cancel each other.\\

Finally, we would like to stress
that the technique we used for deriving the $\delta$-PT2 functional
(which relies on the Legendre--Fenchel transform formalism) 
is expected to be useful also for the development of
{\it
ab initio} 
weight-dependent exchange and correlation DFAs in finite model
systems like electrons on a hypersphere~\cite{PRL09_Loos_2e_hypersphere}.  
Work is currently in progress in this direction.

\section*{Acknowledgments}

This work was funded by the Ecole Doctorale des Sciences Chimiques 222 
(Strasbourg) and the ANR (MCFUNEX project, Grant No. ANR-14-CE06- 0014-01).
The authors are pleased to dedicate this work to Hardy Gross on the
occasion of his 65th birthday.

\section{Authors contributions}

All the authors were equally involved in the preparation of the
manuscript (derivation of equations, implementation of the various
approximations, test calculations and writing).
All the authors have read and approved the final manuscript.

\appendices

\numberwithin{equation}{section}
\section{Appendix: simplified expression for $\left.\frac{\ddroit^2 f^w(\delta)}{\ddroit
\delta^2}\right|_{\delta=0}$}\label{appendix:simplify_d2fwd2delta}
By differentiating Eq.~(\ref{eq:diff_nu_3rd_order_eq}) with respect to
$\nu$ and taking $\nu=\nu(0)=0$ it comes 
\be\label{eq:2nd_deriv_ei_wrt_nu}
&&\left.\dfrac{ \partial^2 e_i(\nu)}{\partial
\nu^2}\right|_{\nu=0}\times\Big[1-u^2+4ue_i(0)-3e_i^2(0)\Big]
\nonumber\\
&&=-2e_i(0),
\ee
since, according to Eq.~(\ref{eq:diff_nu_3rd_order_eq}), $\left.\partial
e_i(\nu)/\partial \nu\right|_{\nu=0}=0$.   
In the particular (symmetric) case $\nu=0$,
Eq.~(\ref{eq:3rd_order_poly_eq_ener}) becomes
\be\label{from_3rd_to_2nd_order_sym}
e_i=u
%\nonumber\\
 \hspace{0.4cm}\mbox{or}\hspace{0.4cm}
%\nonumber\\
e^2_i-ue_i-1=0,
\ee
thus leading to
\be\label{eq:e0_exp}
e_0(0)=\dfrac{u-\sqrt{u^2+4}}{2}
\ee
and
\be\label{eq:e1_exp}
e_1(0)=u.
\ee
In the following, we will use the shorthand notation $e=e_0(0)$ for
convenience. 
Consequently, we obtain from
Eq.~(\ref{eq:2nd_deriv_ei_wrt_nu}) the following explicit expressions, 
\be
\left.\dfrac{ \partial^2 e_0(\nu)}{\partial
\nu^2}\right|_{\nu=0}
=-\dfrac{2e}{1-u^2+4ue-3e^2}
,
\ee
or, equivalently (see Eq.~(\ref{from_3rd_to_2nd_order_sym})),
\be\label{eq:d2e0d2nu}
\left.\dfrac{ \partial^2 e_0(\nu)}{\partial
\nu^2}\right|_{\nu=0}
=\dfrac{2e}{2+u^2-ue}
,
\ee
and
\be\label{eq:d2e1d2nu}
\left.\dfrac{ \partial^2 e_1(\nu)}{\partial
\nu^2}\right|_{\nu=0}=-2u.
\ee
Combining Eqs.~(\ref{eq:d2fwd2delta0}), (\ref{eq:lr_vector_pot}),
(\ref{eq:d2e0d2nu}), and (\ref{eq:d2e1d2nu}) leads to 
\be
\left.\dfrac{\ddroit^2 f^w(\delta)}{\ddroit
\delta^2}\right|_{\delta=0}=
\dfrac{1}{2}
\dfrac{2+u^2-ue}{w\left[e+u\left(2+u^2-ue\right)\right]-e}
.
\ee
Finally, by using the simplified expression,
\be
2+u^2-ue=
1+\left(\frac{u}{2}+\sqrt{1+\left(\frac{u}{2}\right)^2}\right)^2,
\ee
we obtain (see Eq.~(\ref{eq:def_g(u)}))
\be
\dfrac{2+u^2-ue}{e}=-g(u),
\ee
thus leading to the final expression in Eq.~(\ref{eq:final_exp_d2fwd2w}).

\section{Appendix: simplified expressions for 
$\left.\frac{\ddroit f^w(\delta,u)}{\ddroit
u}\right|_{u=0}$ 
and
$\left.\frac{\ddroit^2 f^w(\delta,u)}{\ddroit
u^2}\right|_{u=0}$ 
}\label{appendix:simp_deriv_u}

According to Eq.~(\ref{eq:3rd_order_poly_eq_ener}),
% and (\ref{eq:diff_nu_3rd_order_eq})
the
individual energies 
read as
follows in the non-interacting case ($u=0$),
\be
e_0(\nu,0)&=&-\sqrt{1+\nu^2},
\nonumber\\
e_1(\nu,0)&=&0 
,
\ee
so that
\be
\left.\dfrac{\partial e_0(\nu,u)}{\partial
\nu}\right|_{u=0}&=&
%-\dfrac{2\nu e_0(\nu,0)}{1+\nu^2-3e^2_0(\nu,0)}
%\nonumber\\
%&=&
-\dfrac{\nu}{\sqrt{1+\nu^2}}
,
\nonumber\\
\left.\dfrac{\partial e_1(\nu,u)}{\partial
\nu}\right|_{u=0}
&=&0,
\ee  
and
\be\label{eq:d2ei_dnu2}
\left.\dfrac{\partial^2 e_0(\nu,u)}{\partial
\nu^2}\right|_{u=0}&=&
%-\dfrac{2\nu e_0(\nu,0)}{1+\nu^2-3e^2_0(\nu,0)}
%\nonumber\\
%&=&
-\dfrac{1}{\left(1+\nu^2\right)^{3/2}}
,
\nonumber\\
\left.\dfrac{\partial^2 e_1(\nu,u)}{\partial
\nu^2}\right|_{u=0}
&=&0.
\ee
Moreover, we obtain the following expressions from the differentiation
of Eq.~(\ref{eq:3rd_order_poly_eq_ener}) with respect to $u$ and/or
$\nu$:
\be\label{eq:de0du}
\left.\dfrac{\partial e_0(\nu,u)}{\partial
u}\right|_{u=0}&=&\dfrac{1-2e^2_0(\nu,0)}{1+\nu^2-3e^2_0(\nu,0)}
\nonumber\\
&=&\dfrac{\nu^2+\frac{1}{2}}{\nu^2+1},
\ee
\be\label{eq:de1du}
\left.\dfrac{\partial e_1(\nu,u)}{\partial
u}\right|_{u=0}
&=&\dfrac{1}{\nu^2+1},
\ee
%and
\be\label{eq:d2e0_du2}
&&\left.\dfrac{\partial^2 e_0(\nu,u)}{\partial
u^2}\right|_{u=0}=
\dfrac{
2e_0(\nu,0)}
{1+\nu^2-3e^2_0(\nu,0)}\times
\nonumber\\
&&\left[1+\left.\frac{\partial e_0(\nu,u)}{\partial
u}\right|_{u=0}\left(3\left.\frac{\partial e_0(\nu,u)}{\partial
u}\right|_{u=0}-4\right)\right]
\nonumber\\
&&=-\dfrac{\nu^2+\frac{1}{4}}{\left(1+\nu^2\right)^{5/2}},
%\nonumber\\
%\nonumber\\
\ee

\be\label{eq:d2e1_du2}
&&\left.\dfrac{\partial^2 e_1(\nu,u)}{\partial
u^2}\right|_{u=0}=0,
\ee

\be\label{eq:d2e0_dnudu}
&&\left.\dfrac{\partial^2 e_0(\nu,u)}{\partial\nu\partial
u}\right|_{u=0}=
\dfrac{
2}
{1+\nu^2-3e^2_0(\nu,0)}\times
\nonumber\\
&&\Bigg[e_0(\nu,0)
\left.\dfrac{\partial e_0(\nu,u)}{\partial
\nu}\right|_{u=0}
\left(3\left.\dfrac{\partial e_0(\nu,u)}{\partial
u}\right|_{u=0}-2\right)
\nonumber\\
&&-
\nu
\left.\dfrac{\partial e_0(\nu,u)}{\partial
u}\right|_{u=0}
\Bigg]
\nonumber\\
&&=\dfrac{\nu}{\left(1+\nu^2\right)^2}
,
\ee

\be\label{eq:d2e1_dnudu}
\left.\dfrac{\partial^2 e_1(\nu,u)}{\partial\nu\partial
u}\right|_{u=0}&=&-\dfrac{2\nu
}{1+\nu^2}
\left.\dfrac{\partial e_1(\nu,u)}{\partial
u}\right|_{u=0}
\nonumber\\
&=&-\dfrac{2\nu}{\left(1+\nu^2\right)^2}
.
\ee
Combining Eqs.~(\ref{eq:dfw_over_du}),
(\ref{eq:de0du}) and (\ref{eq:de1du}) leads to
\be
\left.\dfrac{\ddroit f^w(\delta,u)}{\ddroit
u}\right|_{u=0}=\left.\dfrac{1+w+2\nu^2(1-w)}{2(1+\nu^2)}\right|_{\nu=\nu^w(\delta,0)}.\nonumber \\
\ee
By inserting Eq.~(\ref{eq:KS_pot_over_2t}) into the latter equation we
finally recover, as expected, the expression for the exact ensemble Hx energy (see
Eq.~(\ref{eq:ensHx_over_2t}))
per unit of $u$:
\be
\left.\dfrac{\ddroit f^w(\delta,u)}{\ddroit
u}\right|_{u=0}=
e_{\rm Hx}^w(\delta)/u.
\ee
Turning to the ensemble correlation energy, it comes from
Eqs.~(\ref{eq:LR_vec}), (\ref{eq:d2ei_dnu2}), (\ref{eq:d2e0_dnudu}), and
(\ref{eq:d2e1_dnudu}) that  
\be
\left.\dfrac{\partial {\nu}^w(\delta,u)}{\partial u}\right|_{u=0}
=
\left.\dfrac{\nu(1-3w)}{(1-w)\sqrt{1+\nu^2}}
\right|_{\nu=\nu^w(\delta,0)},
\ee
which, according to Eqs.~(\ref{eq:d2fw_du2_0}), (\ref{eq:d2e0_du2}), and (\ref{eq:d2e1_du2}) leads to 
\be
\left.\dfrac{\ddroit^2 f^w(\delta,u)}{\ddroit
u^2}\right|_{u=0}
&=&-\Bigg(\dfrac{(1-w)}{4(1+\nu^2)^{5/2}}\times
\\
&&
%\left.
\left[1+4\nu^2-\dfrac{4\nu^2(1-3w)^2}{(1-w)^2}\right]
%\right|
\Bigg)_{\nu=\nu^w(\delta,0)} \nonumber
.\ee
Finally, by using the following relations (see 
Eq.~(\ref{eq:KS_pot_over_2t})),
\be
\dfrac{1}{1+\big[\nu^w(\delta,0)\Big]^2}&=&\dfrac{(1-w)^2-\delta^2}{(1-w)^2}
,
\nonumber\\
\dfrac{1+4\big[\nu^w(\delta,0)\Big]^2}{1+\big[\nu^w(\delta,0)\Big]^2}&=&
1+\dfrac{3\delta^2}{(1-w)^2},
\nonumber\\
%\big[\nu^w(\delta,0)\Big]^2&=&\dfrac{\delta^2}{(1-w)^2-\delta^2},
\dfrac{\big[\nu^w(\delta,0)\Big]^2}{1+\big[\nu^w(\delta,0)\Big]^2}&=&\dfrac{\delta^2}{(1-w)^2},
\ee
we recover the expression in Eq.~(\ref{eq:Taylor_u_second_order}). 

\section{Appendix: correlation energy at the border of the $v$-representability
domain}\label{appendix:Ec_border}

As readily seen from Eq.~(\ref{eq:non_int_vrep_cond}), at the border of
the non-interacting $v$-representability
domain, the density is such that $\vert n-1\vert=1-w$ or, equivalently,
\be
n=1\pm (1-w).
\ee 
When $\vert\Delta v\vert/t\rightarrow +\infty$ and $\vert\Delta
v\vert>U$, the
ground- and first-excited state energies read as follows, according to
Eq.~(\ref{eq:3rdorder_energy_prb}),
\be
E_0(\Delta v)&=&U-\vert\Delta v\vert,
\nonumber\\
E_1(\Delta v)&=&0,
\ee
and, consequently (see Eq.~(\ref{eq:diff_nu_3rd_order_eq})),
\be
\dfrac{\partial E_0(\Delta v)}{\partial \Delta v}&=&-\dfrac{\Delta
v}{\vert\Delta v\vert},
\nonumber\\
\dfrac{\partial E_1(\Delta v)}{\partial \Delta v}&=&0.
\ee
Thus we conclude that the stationarity condition in
Eq.~(\ref{eq:stationarity_cond_LF}) is fulfilled for $\delta=n-1=\pm (1-w)$ when $\vert\Delta
v\vert/t\rightarrow +\infty$ and $\Delta v/(n-1)$ is {\it positive}. The
resulting ensemble Legendre--Fenchel transform (see Eq.~\ref{universalfunctionalHD}) reads 
\be\label{eq:Fw_border}
&&(1-w)\Big(U-\vert\Delta v\vert\Big)\pm\Delta v\Big(1-w\Big) 
\nonumber\\
&&\underset{\Delta v\rightarrow \pm\infty}{\longrightarrow}
F^w\Big(1\pm (1-w)\Big)
=U(1-w).
\ee
Since, according to Eq.~(\ref{eq:noninteractingenergyhubbarddimer}),
\be
T^w_{\rm s}\Big(1\pm (1-w)\Big)=0,
\ee
it comes from Eqs.~(\ref{eq:EEXX_fun_HD}) and (\ref{eq:Fw_border}),
\be
&&\left.\Big[F^w(n)-T^w_{\rm s}(n)\Big]\right|_{n=1\pm (1-w)}
\nonumber\\
&&=\left.\Big[E_{\rm H}(n)+E^w_{\rm
x}(n)\Big]\right|_{n=1\pm (1-w)}
,\ee  
or, equivalently,
\be
E^w_{\rm
c}\Big(1\pm (1-w)\Big)=0.
\ee
\section{Appendix: EEXX ensemble energy minimization in the symmetric case}\label{appendix:eexx_min}

For $\Delta v_{\rm ext} = 0$, the minimization of the (approximate) EEXX 
ensemble energy leads to the following equation:
\be
(n-1)\left[\frac{U(1-3w)}{(1-w)^2}+\frac{2t}{\sqrt{(1-w)^2-(1-n)^2)}} \right] = 0. \nonumber \\
\ee
After factoring out the obvious solution $n=1$, we are left with a quadratic equation. The discriminant reads
\be
\Delta = 4U^2(1-3w)^2(w-1)^2\left[U^2(1-3w)^2 -4t^2(w-1)^2\right] \nonumber \\
\ee
and is zero for $U=0$ and the critical value
\be
U_{\rm crit} = \frac{2t(1 - w)}{3w - 1}.
\ee
The second derivative of the ensemble energy functional with respect to $n$ contains all the 
information about the convexity:
\be
\frac{\ddroit^2 E^{w}_{\rm{EEXX}}(n)}{\ddroit n^2}=
\frac{U(1-3w)}{(1 - w)^2} +
\frac{2t (1-w)^2}{\left[(1-w)^2 - (1-n)^2\right]^{3/2}}.
\nonumber \\
\label{eq:d2E_wdn2}
\ee
For $U\leq U_{\rm crit}$, the EEXX ensemble energy is strictly convex and has exactly one unique 
global minimum ($n=1$) whereas for $U>U_{\rm crit}$ the quadratic 
equation possesses two solutions,
\be
n = 1\pm\frac{\sqrt{\Delta}}{2U^2(3w - 1)^2},
\ee
which leads to two degenerate minima. The other solution $n=1$ is a maximum in 
this case (see Eq.~(\ref{eq:d2E_wdn2}) and the top panel of 
Fig.~\ref{fig:ensemble_energy_profile}). We notice that for any $w\leq 1/3$, the 
ensemble energy within the EEXX approximation has always one global
minimum, 
independently of $U$. In the case of
equi--ensembles ($w=1/2$), there is one unique solution as long as $U\leq 2t$.

%%%%%%%%%%%%%%%%%%%%%%%%%

%
% BibTeX users please use
 \bibliographystyle{epj}
% \bibliography{biblio}

%%%%%% Refs. from the .bbl file

\newcommand{\Aa}[0]{Aa}

%%%%%%%%%%%%%%%%%%%%%%%%%%%%%%%

%\begin{thebibliography}{}
%Author, \textit{Book title} (Publisher, place year) page numbers
% etc
%\end{thebibliography}

\end{document}